\documentclass{JINST}
\usepackage{graphicx}
\usepackage{url}
\let\ifpdf\relax

\newcommand{\Het}{$^3$He }
\newcommand{\Cf}{$^{252}$Cf }

\newcommand{\Cs}{$^{137}$Cs }
\newcommand{\Co}{$^{60}$Co }

\newcommand{\nistFN}{\footnote{Certain commercial equipment, instruments, or materials are identified in this paper in order to specify the experimental procedure adequately. Such identification is not intended to imply recommendation or endorsement by the National Institute of Standards and Technology, nor is it intended to imply that the materials or equipment identified are necessarily the best available for the purpose.} }

\title{Development and Characterization of a High Sensitivity Segmented Fast Neutron Spectrometer (FaNS-2)}

\author{T.J. Langford$^a$\thanks{Corresponding author, thomas.langford@yale.edu}, E.J. Beise$^b$, H. Breuer$^b$, C.R. Heimbach$^c$, G. Ji$^b$\thanks{Current address: Department of Physics, Harvard University, Cambridge, MA 02138 USA}, and J.S. Nico$^c$\\
\llap{$^a$} Wright Laboratory\\
  Yale University \\
  New Haven, CT 06511 USA\\
\llap{$^b$} Department of Physics \\
  University of Maryland,\\
  College Park, MD 20742 USA\\
\llap{$^c$} Physical Measurement Laboratory \\
National Institute of Standards and Technology,\\
  Gaithersburg, MD 20899 USA\\
}

\abstract{
We present the development of a segmented fast neutron spectrometer (FaNS-2) based upon plastic scintillator and \Het proportional counters. 
It was designed to measure both the flux and spectrum of fast neutrons in the energy range of few MeV to 1~GeV. 
FaNS-2 utilizes capture-gated spectroscopy to identify neutron events and reject backgrounds.
Neutrons deposit energy in the plastic scintillator before capturing on a \Het nucleus in the proportional counters. 
Segmentation improves neutron energy reconstruction while the large volume of scintillator increases sensitivity to low neutron fluxes. 
A main goal of its design is to study comparatively low neutron fluxes, such as cosmogenic neutrons at the Earth's surface, in an underground environment, or from low-activity neutron sources. 
In this paper, we present details of its design and construction as well as its characterization with a calibrated \Cf source and monoenergetic neutron fields of 2.5~MeV and 14~MeV.
Detected monoenergetic neutron spectra are unfolded using a Singular Value Decomposition method, demonstrating a 5\% energy resolution at 14~MeV.
Finally, we discuss plans for measuring the surface and underground cosmogenic neutron spectra with FaNS-2.
}


\begin{document}

\section{Introduction}

A wide range of experiments operate detectors that are susceptible to fast neutron-induced backgrounds~\cite{Formaggio04}. 
To avoid these backgrounds, many collaborations must operate their experiments in deep underground laboratories where cosmic rays are shielded by earthen overburden~\cite{Gaitskell04,Angloher2012,Aprile2012,Agnese2014,Akerib2014,Abdurashitov2009, McKinsey05, Aharmim07,Elliott02,Arnaboldi2004, Aalseth2011,Ackerman2011,Agostini2013}. 
However, there are many experiments that cannot operate in such labs, such as short baseline reactor neutrino oscillation searches~\cite{ashenfelter2013prospect, Danilov:2013caa, Serebrov:2012sq, Pequignot2015126, Lane:2015alq}.
Experiments typically address these backgrounds with passive or active shielding, designed for their specific neutron environment. 
Neutrons with energy below 10~MeV are readily shielded using conventional hydrogenous materials. 
However, neutrons above 10~MeV from cosmic ray interactions are deeply penetrating and cannot be effectively shielded~\cite{Mei2006,Carson2004}. 
Understanding the neutron spectrum and flux in this energy range has great impact on the sensitivity of operating and proposed experiments.

Many different detector technologies have been developed to study fast neutrons in various energy ranges, including Bonner sphere arrays~\cite{Bramblett1960,Alevra2002,Gordon2004,Goldhagen2002, Park2013302}, liquid scintillator detectors~\cite{Boehm2000,Nakao2001,bellini2011muon,Zhang2013138,zhang2014LSatSoudan}, crystalline detectors~\cite{1960BroekStilbene,Osipenko:2015ysa,Osipenko:2015dta,Caiffi:2015zsa}, and pressurized helium detectors~\cite{Manolopoulou2006371}. 
Capture-gated spectroscopy is a technique that uses delayed neutron captures to identify neutron interactions in an active medium~\cite{Czirr1989,Abdurashitov2002a,Czirr2002,Normand2002,Fisher2011,Bass2012}.
Capture-gated detectors based on loaded liquid scintillator have been deployed in various environments~\cite{ALEKSAN1989203,Bonardi2010225}.
However, these detectors can suffer from stability issues because liquid scintillator changes density with temperature and the light yield can degrade without a nitrogen cover-gas.
Recently, special formulae of loaded plastic scintillator have been shown to demonstrate pulse shape discrimination between neutron and gamma events~\cite{cherepy2015bismuth, zaitseva2013pulse}. These new materials are in their early stages of development and may not be ready for large-volume deployments for some time.

A heterogenous array of \Het proportional counters and plastic scintillator provides a rugged and stable detector that can be deployed in a wide range of environments.
In these detectors, fast neutrons thermalize in the plastic scintillator and their energy is converted to light detected by photomultiplier tubes. 
The resulting thermal neutrons diffuse through the plastic and then are detected by one of the \Het proportional counters. 
Neutron diffusion introduces a time delay between the neutron recoil and capture by a characteristic time distribution related to the distance between scatter and capture locations.
For many detector configurations and geometries this distribution can be well-described by an exponential decay.
This delayed coincidence is a powerful tag for neutron interactions and eliminates the majority of non-neutron events.
The neutron capture is in a detector that is nearly gamma-insensitive, which allows for operation in an environment where gamma-rays far out number neutrons.
Correlated coincidences follow a distinct time structure that can be characterized by an exponential function.
Uncorrelated coincidences are uniform in time and can be subtracted in analysis, allowing for further removal of non-neutron events.

Previous large-volume capture-gated detectors have typically consisted of single scintillation volumes to reduce detector complexity~\cite{Bonardi2010225, Abdurashitov2000, Abdurashitov2002, Abdurashitov2006}.
However, this approach limits the possible detector resolution through reduced light collection efficiency and scintillator nonlinearity effects.
Fast neutrons thermalize via multiple proton recoils, which in organic scintillator have a known nonlinear relation between deposited energy and emitted light.
By segmenting the scintillator volume and separating individual recoils, it is possible to undo this nonlinearity for each recoil and reconstruct the true deposited energy~\cite{Bowden2009}.
The optimal segment size depends on the energy range under study, as higher energy neutrons travel longer distances between scatters.
However, fine segmentation must be balanced with light collection efficiency and complexity of electronics. 
Simulations should be performed to determine the optimum segmentation for each detector and its application.

Neutron capture-gated spectroscopy with \Het proportional counters with plastic scintillator was previously demonstrated with a 15-liter active volume detector, FaNS-1~\cite{2015Langford}. 
This detector was characterized with known neutron fields and then operated remotely at the Kimballton Underground Research Facility (KURF)~\cite{Finnerty2010,2014APSKURF}.
There it measured the flux and energy spectrum of neutrons from local radioactivity in the rock and surrounding environment. 
FaNS-1 was subsequently used to characterize backgrounds in research reactor environments~\cite{ashenfelter2015background}. 

In this work we present a new fast neutron spectrometer (FaNS-2) that was developed combining capture-gated spectroscopy with a large-volume segmented detector.
FaNS-2 was designed to measure the energy spectrum and flux of cosmogenic fast neutrons at the earth's surface and shallow underground laboratories. 
This requires a high-efficiency detector that has sensitivity to a broad range of energies. 
Precise energy reconstruction is also needed to observe structure in the neutron energy spectrum from interactions in the atmosphere.
In this paper we detail the design and characterization of the FaNS-2 detector.
Section~\ref{detectorDesign} discusses the simulation and construction of the detector.
In Section~\ref{analysis}, the analysis methodology is presented for each detector subsystem as well as subtraction of uncorrelated backgrounds.
The response of the detector to gamma, \Cf, and monoenergetic neutron sources is presented in Section~\ref{results}, along with a discussion of the impact of segmentation on energy reconstruction and a demonstration of spectral unfolding via Singular Value Decomposition.
Directional detection with FaNS-2 is presented, including a brief discussion of detection of low-activity neutron sources.
Finally, we discuss how the detector will be used to measure the spectrum and flux of cosmogenic neutrons in a surface and shallow underground environment.

\section{The FaNS-2 detector}
\label{detectorDesign}
\subsection{Simulation of detector design}

The FaNS-1 detector functions well but its design, in particular the number, size, and arrangement of scintillator and \Het proportional counters, was not the result of any optimization studies. 
With FaNS-2, however, Monte Carlo simulations were performed using MCNPX~\cite{hendricks2008mcnpx} to study the response of different designs to gamma and monoenergetic neutron sources. 
Various geometries were simulated, modifying the number of \Het detectors and the size and arrangement of scintillator segments.
The simulated sensitivity to the broad-energy cosmogenic fast neutron spectrum (extending 1~MeV to many GeV, Ref.~\cite{Gordon2004}) of each configuration was used to optimize the response to high-energy neutrons while preserving energy resolution at low energy. 
One of the lessons learned from the various geometries is the benefit of separating the \Het detectors from each other.
Ensuring that no \Het detector shadows any other maximizes the neutron capture efficiency per \Het detector. 
The MCNP simulations also indicated that a more densely packed array increases the total capture efficiency, while a symmetric array provides a more uniform efficiency to neutrons regardless of incident direction. 
FaNS-2 is intended to measure neutrons from a variety of sources, so a uniform response is desirable.  
The simulation results influenced the final design, shown in Figure~\ref{fig:FaNS2diagram}, of the individual detector elements as well as optimizing their arrangement in the detector array.

Optical simulations were performed with a modified Guide7 package~\cite{massam1976guide} to determine the uniformity of the light collection along the scintillator segment. 
A single scintillator bar was modeled in the Monte Carlo simulation, including the material properties of the scintillator, a reflective wrapping surrounding the segment, and light guides coupling to optically sensitive surfaces the same size as the PMT photocathode area.
The total light collection efficiency and uniformity was compared between two light guide shapes: a straight cylinder matched to the PMT diameter, and a tapered cone that transitions between the rectangular cross-section of the scintillator to the PMT diameter. 
While the latter produced higher light collection efficiency than the cylinder, it also introduced larger geometric nonuniformities.
For this reason, cylindrical light guides were attached to the center of each 9$\times$9~cm$^2$ face of the scintillator bars with optical cement.  

\subsection{Detector design and construction}
\label{sec:detectorDesign}
The FaNS-2 detector is an array of 16 EJ-200\nistFN plastic scintillator bars and 21 \Het proportional counters.
The scintillator bars are $9\times9\times56$~cm$^3$ and have 5~cm diameter, 9.6~cm long UVT acrylic light guides on either end attached with optical cement, for a total active volume of 72.5~liters. 
The surfaces of the scintillator segments are diamond-tool finished to improve light propagation via total-internal reflection. 
Additionally, each scintillator segment is loosely wrapped with aluminized mylar and a black vinyl light-tight cover (1.5~mm thick) to further increase light collection efficiency and minimize cross-talk between segments.
Two 5~cm Phillips XP2262B photomultiplier tubes are coupled to the light guides via 5~cm diameter, 3~mm thick EJ-560 silicone rubber optical interface pads. 
Double ended readout of the scintillator segments enhances the light collection uniformity along the scintillator segment.

The \Het detectors are GE-Reuter Stokes model RS-P4-0819-103 with 2.5~cm diameter and 46~cm active length.
The \Het partial pressure in the counters was 404~kPa (4~atm) with a buffer gas consisting of 111~kPa (1.1~atm) of krypton to enhance the stopping power of the gas mixture. 
To benchmark the simulated sensitivity of the \Het proportional counters, raw neutron capture rates in data and Monte Carlo were compared in three \Het detectors with a \Cf source at three distances.
The MCNP simulation was found to systematically over-estimate the count rate by 16\% compared to data. 
Therefore, a (84$\pm$10)\% efficiency is applied to MCNP simulations to account for the discrepancy~\cite{2013LangfordThesis}.
Alpha decays from trace radioactivity in the wall material have previously been identified as an important background for \Het detectors~\cite{Cox2004,Amsbaugh2007}. 
To address these backgrounds, approximately 100 \Het detectors were surveyed to identify those with low alpha activity. 
Selecting those detectors with low alpha emitting content reduces backgrounds for future operation of FaNS-2 in low-neutron environments. 

\begin{figure}[h]
\begin{center}
\includegraphics[width=.32\textwidth]{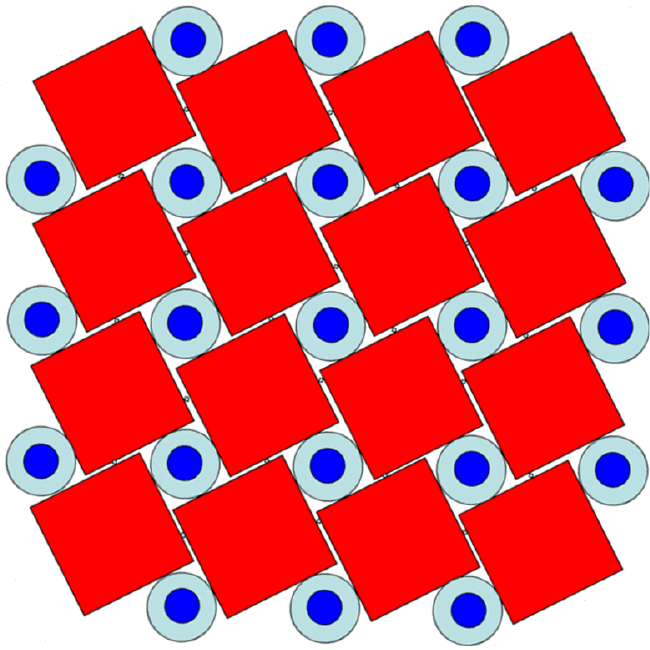}~\hspace{2mm}
\includegraphics[width=.6\textwidth]{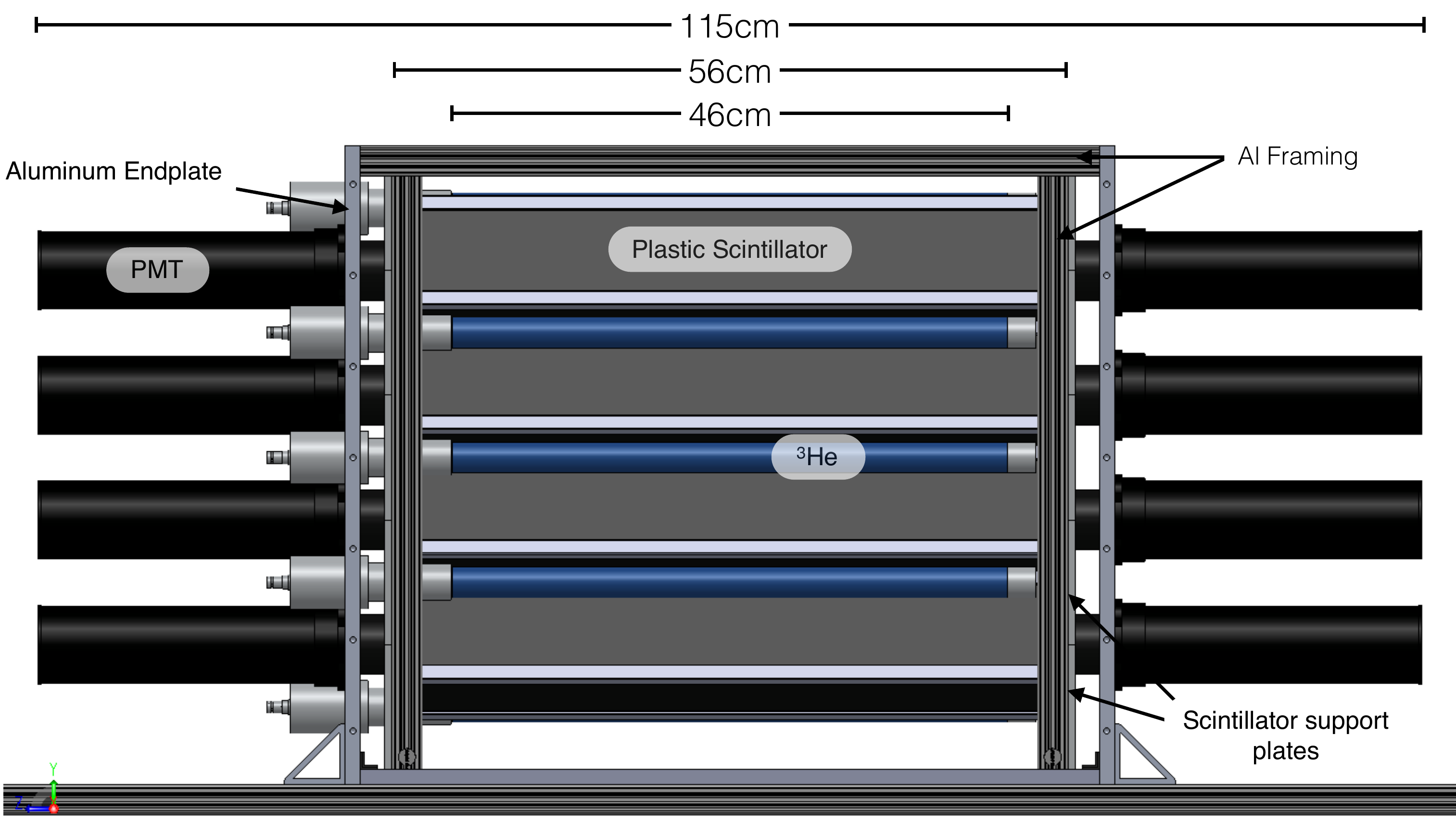}
\caption{Left: MCNP model of the scintillator (red squares) and \Het detector layout (dark blue circles), including the space required for \Het mechanically mounting (light blue circles). Right: A CAD drawing of the assembled FaNS-2 detector array, including PMT housings and aluminum support structure but omitting the outer aluminum side panels. The active length of the scintillator segments is 56~cm, while the active \Het region is 46~cm and highlighted in blue. The full length of the detector assembly, excluding an aluminum frame (not pictured) that protects the PMTs, is 115~cm.}
\label{fig:FaNS2diagram}
\end{center}
\end{figure}

The arrangement of the detector array is designed to maximize the scintillator coverage of the \Het  detectors while maintaining symmetry and dense packing of the elements.
Approximately ten detector packing schemes were modeled in MCNP to determine their overall efficiency to a broad range of neutron energies.
Figure~\ref{fig:FaNS2diagram} shows a schematic of the selected design.
The mounting hardware of the \Het detectors determined the size of the opening created by the scintillator segment tilt.
A CAD rendering of the full detector package is shown in the right panel of Figure~\ref{fig:FaNS2diagram}.

The detector elements are supported by thin aluminum angle brackets that span between two 0.6~cm thick aluminum scintillator support plates mounted to vertical aluminum framing pieces. 
The angle brackets secure each scintillator segment with rubber inserts that minimize vibrations and preserve detector positioning.  
Photomultiplier tubes are housed in spring-loaded light-tight enclosures mounted to 1.2~cm thick outer aluminum endplates that apply pressure between the PMT and optical interface pads, maintaining optical coupling. 
The endplates are connected to each other via aluminum framing to ensure mechanical stability of the assembly.
The \Het detectors are supported by aluminum collars that mount into one of the outer aluminum endplates with o-rings to maintain light-tightness and bayonet pins to hold them in place. 
The full volume of the detector is surrounded by 3~mm thick sheets of boron-loaded silicone rubber (Shieldwerx model SWX-238) to shield from thermal neutrons~\cite{boroflex}. 
The detector assembly is supported by a 1.2~cm thick aluminum baseplate mounted to an aluminum frame.
Thin aluminum sheets seal against the outer endplates plates via black foam rubber to ensure a light-tight environment for the scintillator bars. 
A more detailed description of the detector construction and assembly can be found in Ref.~\cite{2013LangfordThesis}.

After construction, a detailed MCNP model of FaNS-2 was completed that includes the light guides, aluminum enclosure, and inner support structure.
This model was then used for simulations of the detector response to various neutron sources, including \Cf and monoenergetic neutrons from DD and DT interactions.
The aluminum enclosure and support structure was found to contribute very little to neutron interactions within the detector. 
The light collection efficiency is assumed to be uniform throughout the scintillator volume, and photo-statistics and scintillator nonlinearity are applied on an event-by-event basis. 
A detector response function was generated through simulations of monoenergetic neutrons with a wide range of incident energy.
This is used to unfold detected neutron spectra, detailed in Section~\ref{sec:unfolding}. 

\subsection{Electronics and data acquisition}
\label{sec:electronics}

\begin{figure}[h]
\begin{center}
\includegraphics[width=0.85\textwidth]{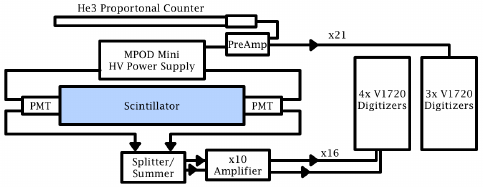}
\caption{A schematic of the data acquisition and electronics for FaNS-2. More detail is found in the text and in Ref.~\cite{2013LangfordThesis}.}
\label{fig:FaNS2electronics}
\end{center}
\end{figure}

\begin{figure}[htbp]
\begin{center}
\includegraphics[width=.48\textwidth, height=1.8in]{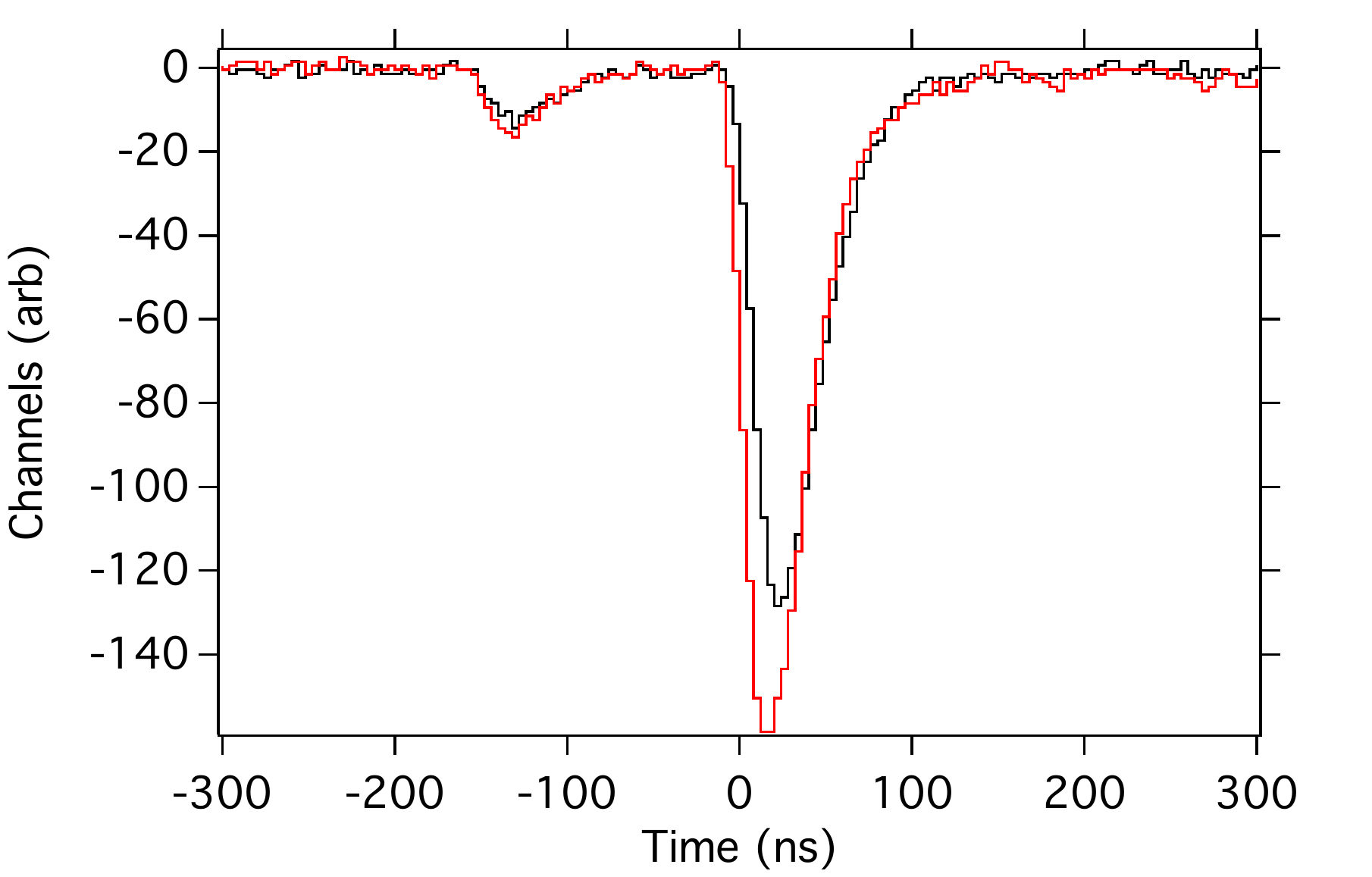}~
\includegraphics[width=.48\textwidth, height=1.8in]{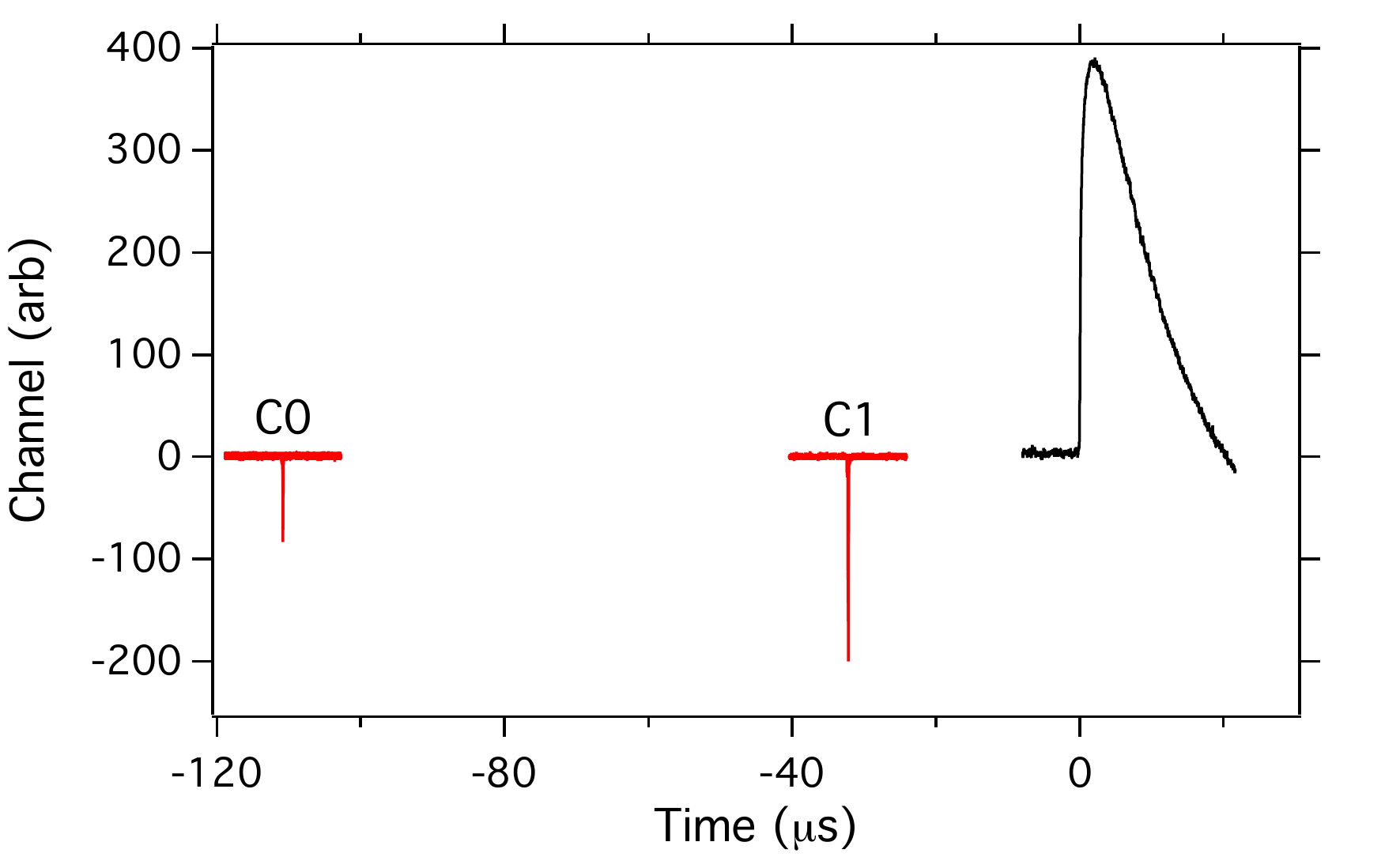}
\caption{Left: An example trace of two PMT signals from a single scintillator bar. The attenuated pre-pulse can be clearly seen. This pre-pulse has 1/10th of the full integral of the full pulse that follows, yielding an extended dynamic range. Right: An example acquisition window from one trigger with one \Het signal (black) and two clusters of multiple PMT signals (red, labeled C0 and C1). These are stored in binary format for offline analysis.}
\label{fig:traces}
\end{center}
\end{figure}

A schematic of the FaNS-2 electronics system is shown in Figure~\ref{fig:FaNS2electronics}.
Both the PMT and \Het proportional counters are supplied with high voltage from an MPod Mini crate outfitted with ISEG high voltage cards.
Signals from the 32 PMTs and 21 \Het proportional counters are read out by seven CAEN V1720 waveform digitizers (8~channel, 12~bit digitization resolution, 250~MSamples/s) operated synchronously. 
The digitizer is triggered internally by any \Het detector, which is propagated to each digitizer simultaneously. 
Any scintillator signal over threshold in the acquisition window (-200~$\mu$s, 600~$\mu$s) is written to disk for off-line analysis. 
The large acquisition window combined with the fast sampling frequency can yield prohibitively large data rates. 
However, by using the on-board Zero Suppression algorithm, a small snippet of data surrounding a signal can be stored without losing the signal location in the acquisition window.
Figure~\ref{fig:traces} shows an example of a digitized acquisition window with one \Het signal and multiple clusters of PMT signals.

The PMT signals are routed through a custom electronics module that asymmetrically splits each signal (10\% and 90\%), delays one branch by 150~ns, and re-sums the two branches~\cite{Breuer13}.
This results in the signal shown in Figure~\ref{fig:traces}, where an attenuated copy of the signal precedes the full signal. 
If the full branch of the signal saturates the electronics, the attenuated branch can be used instead. 
The dynamic range of the scintillator energy response is extended by a factor of ten while maintaining good digitization resolution for small signals.
The module also increases the PMT signal FWHM from approximately 10~ns to 50~ns to reduce effects caused by the finite sampling frequency of the digitizers. 
The PMT voltage dividers were specially designed to produce a highly linear response ($<10\%$ deviation up to 10~V).
PMT linearity was measured with a picosecond laser that passed through various neutral density filters and a diffuser to illuminate the full PMT face.
A further discussion can be found in Ref.~\cite{2013LangfordThesis}.

The \Het proportional counters are powered through CAEN A1422 four-channel charge-sensitive preamplifier modules.
While typically these preamplifier signals would be passed through a shaping amplifier, we choose to directly digitize the raw signals, an example of which can be seen in Figure~\ref{fig:traces}
The identification of particle type based on risetime and energy analysis is discussed in more detail in Section~\ref{sec:analysisHet}. 

\section{Analysis}
\label{analysis}

\subsection{Nonlinear light response in plastic scintillator}
\label{subsec:lightResponse}
The light response of organic scintillator to heavy charged particles (e.g. recoil protons) is nonlinear in this energy regime. 
Therefore, the deposited energy for each segment is first converted from light (MeV$_{\textrm{ee}}$) to incident energy (MeV) using the light response function. 
A common functional form of this nonlinear response, detailed in Refs~\cite{Birks1951, Craun1970, O'Rielly1996}, is

\begin{equation}
\label{eqn:Craun}
dL/dx = S (dE/dx)\left[1+kB(dE/dx) + C(dE/dx)^2\right]^{-1},\\
\end{equation}

\noindent where $dL/dx$ is the light produced in path length $dx$, $E$ is the particle's energy, $dE/dx$ is the specific energy loss of the particle at the specific energy, and \textit{kB} and \textit{C} are tunable Birks parameters~\cite{Craun1970}. 
The total light is the result of summing over the full range of the particle. 
The specific energy loss of protons in EJ-200 approximated by that of polyvinyltoluene scintillator in the pSTAR database~\cite{Berger2011}. 

Birks parameters are known to vary between scintillator type and manufacturer~\cite{O'Rielly1996}. 
Therefore, the light response function for the EJ-200 in this work was determined through a \Cf neutron time-of-flight (nTOF) setup, as detailed in Ref~\cite{Enqvist2013}. 
Spontaneous fissions from \Cf source emit neutrons along with multiple gamma rays, which can be detected to identify the precise time of fission. 
This can be used in an nTOF apparatus to kinematically determine the energy of an incident neutron based on the flight time to a scintillator segment. 
Two measurements were performed with a FaNS-2 scintillator segment placed either 1~m or 2~m from the tagged \Cf source.
The two distances allow for a cross-check on the method, and they are in agreement with each other.

\begin{figure}[h]
\begin{center}
\includegraphics[width=.55\textwidth]{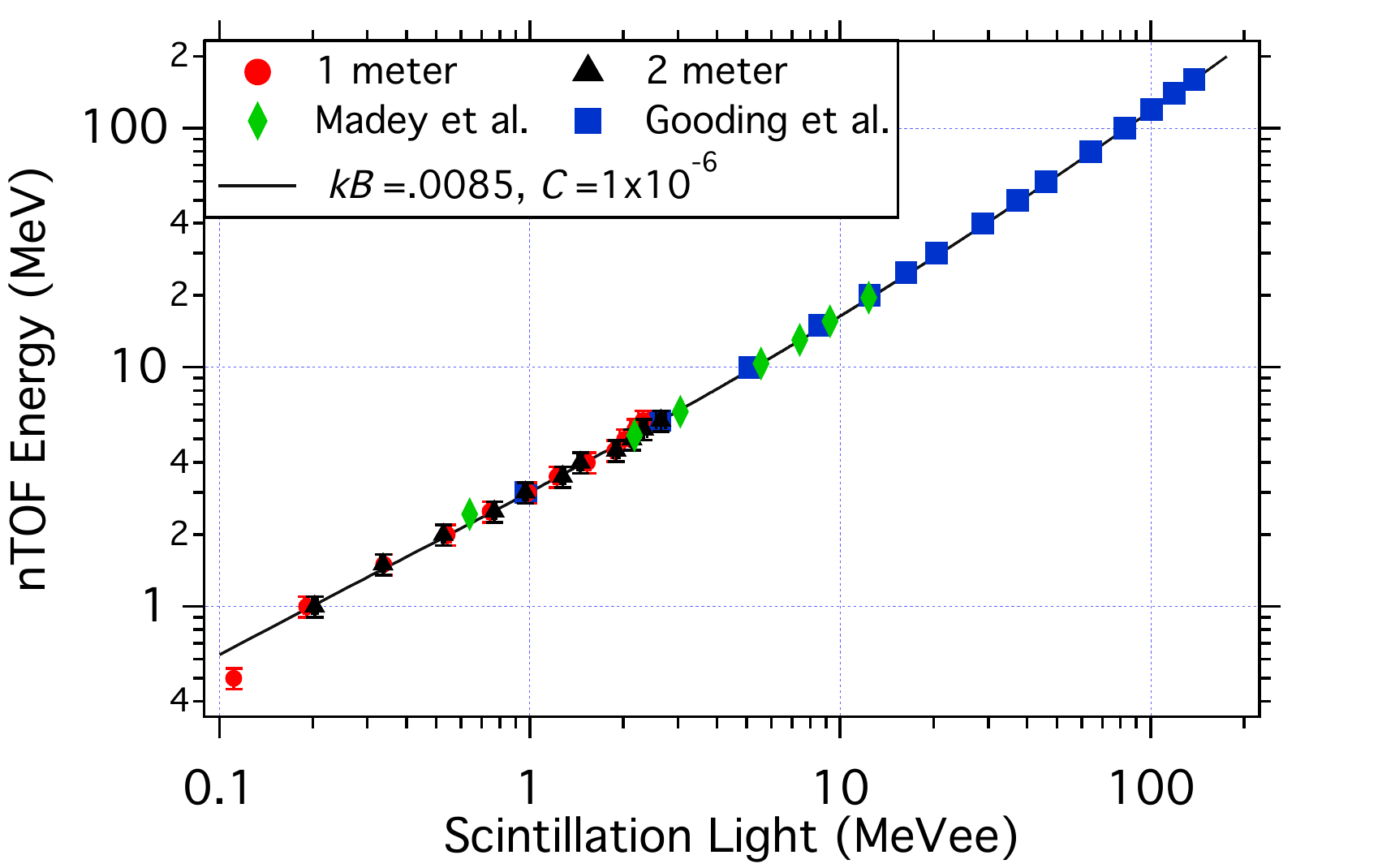}
\caption{The measured response of EJ-200 plastic scintillator to a neutron time-of-flight \Cf source, at distances of 1~m and 2~m, compared to a Birks functional form with parameters: $kB = 8.5\times10^{-3}$~g/cm$^2$/MeV and $C = 1\times10^{-6}$~(g/cm$^2$/MeV)$^2$. Also shown are data points from previous measurements of a similar plastic scintillator, NE-102, from Ref~\cite{Madey1978,gooding1960}.}
\label{fig:nTOF_LR}
\end{center}
\end{figure}

In analysis, slices were made in nTOF energy and the detected energy spectrum for each slice was fitted with a Hill equation to determine the half-height of the distribution~\cite{IgorHillEq}.
These half-heights represent neutrons that deposit all their energy in a single scatter. 
Therefore, by fitting the half-height locations as a function of the nTOF energy, the nonlinear light response can be constructed.
The Birks parameters were then empirically determined to match the nTOF measurement.
Figure~\ref{fig:nTOF_LR} shows the measured light response compared to the Birks functional form with $kB = 8.5\times10^{-3}$~g/cm$^2$/MeV and $C = 1\times10^{-6}$~(g/cm$^2$/MeV)$^2$.
There are a few measurements in this energy range collected with a similar plastic scintillator, NE-102~\cite{Madey1978,gooding1960}. 
These measurements make use of accelerators and are able to extend to much higher energies, though they only have a few data points below 5~MeV.
The nTOF results from this work agree well with the previous measurements and the Birks parameterization describes the low and high energy data well.

\subsection{Scintillator analysis}
\label{subsec:analysisScint}
Light emitted by charged particles in the scintillator is detected by two photomultiplier tubes on each segment. 
The time of the pulse is determined by the half-height of the leading edge of the signal.  
As described in Section~\ref{sec:electronics}, the scintillator signals are split into two branches, one of which is attenuated.
The exact attenuation factor is determined for each PMT by averaging over many pulses that are large enough to have sufficient attenuated signals while not saturating the full branch. 
The pulse integrals for both the full and attenuated branches are calculated and the attenuation factor is corrected for.  
Using gamma calibration data, these integrals are converted into electron-equivalent energies (MeV$_{\textrm{ee}}$) before the two PMTs on each segment are combined.
The electron-equivalent energies are then converted into deposited neutron energy via the nonlinear light response discussed above.
Clusters of scintillator signals from a single particle interacting in multiple segments are defined as those occurring within a 100~ns window of each other.   
The total energy of an event is determined by summing the deposited neutron energies from each segment included in the cluster.
By converting the electron-equivalent energy into deposited neutron energy for each segment individually, FaNS-2 can reproduce the true neutron energy.
This is demonstrated in Sec.~\ref{sec:segmentation} using monoenergetic neutron sources.

\subsection{\Het proportional counter analysis}
\label{sec:analysisHet}

\begin{figure}[h]
\begin{center}
\includegraphics[width=.48\textwidth]{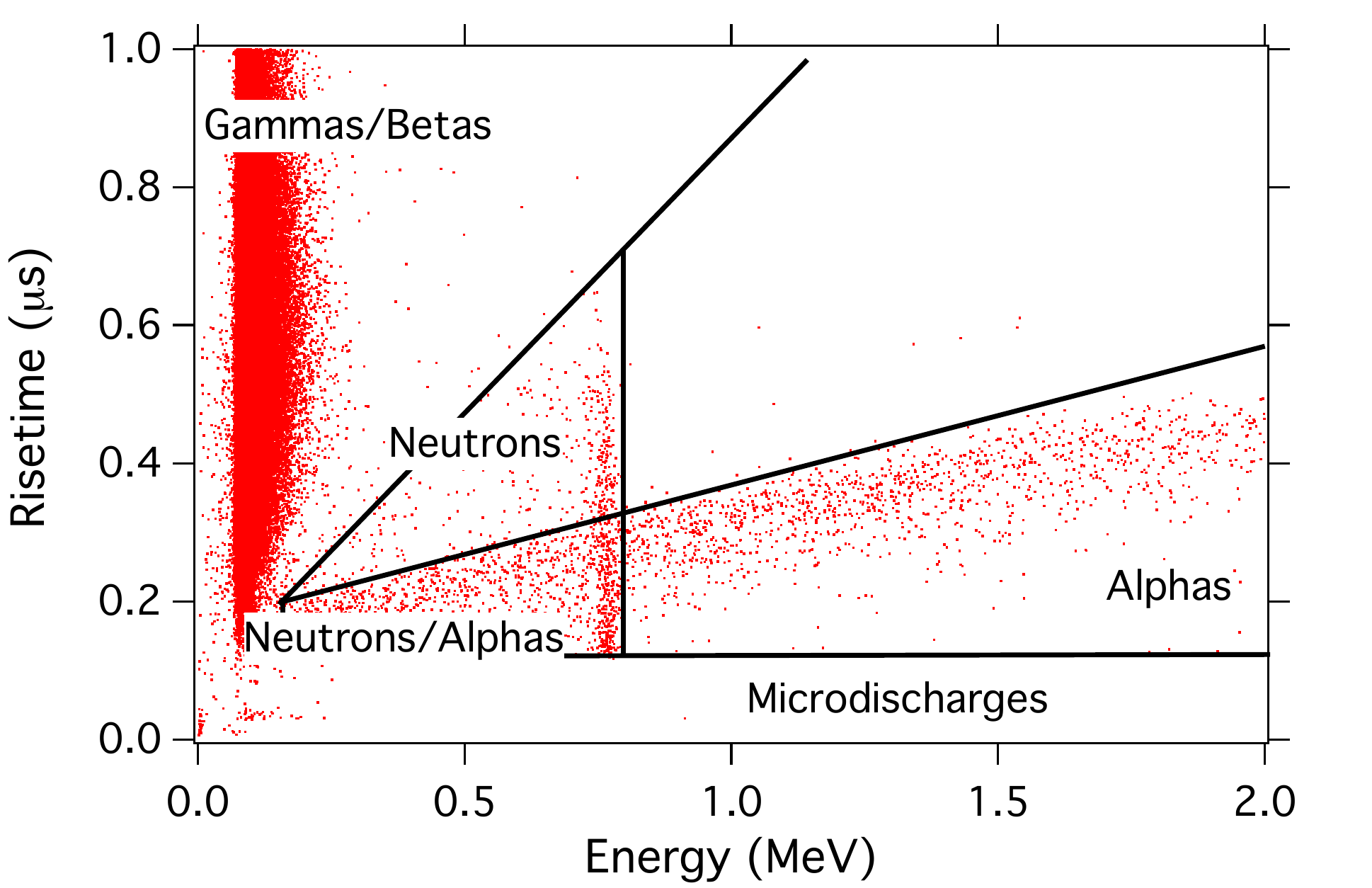}~
\includegraphics[width=.48\textwidth, height=1.8in]{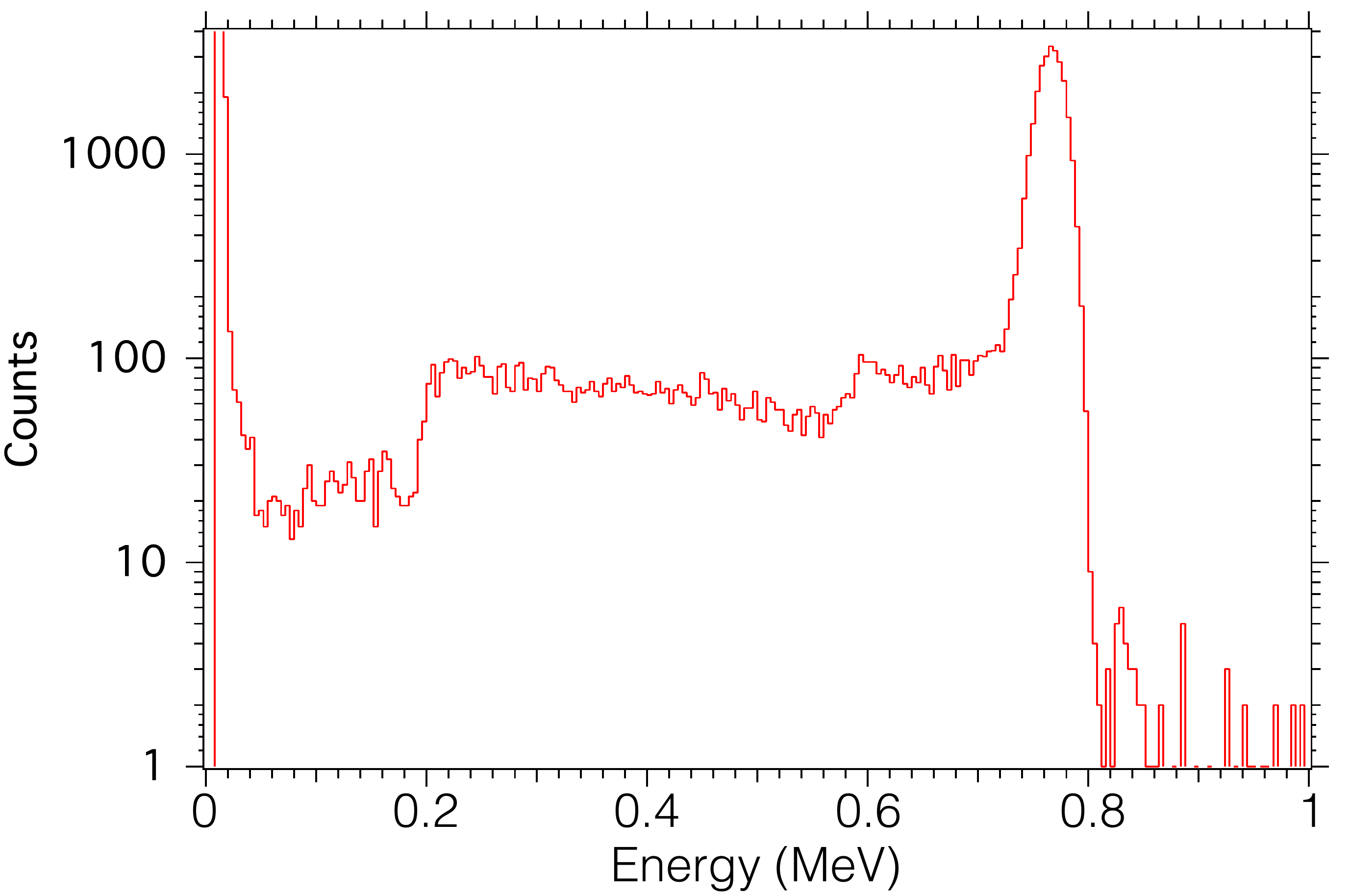}
\caption{Left: A scatter plot of the \Het risetime versus energy with different particle interactions labeled. Clear populations of different particle interactions can be identified. Right: The energy spectrum from the same \Het detector when exposed to a moderated \Cf neutron source. The two ``wall-effect'' edges can be seen at 0.2~MeV and 0.6~MeV. More details are available in Ref.~\cite{2013LangfordHe3}.}
\label{fig:HePlots}
\end{center}
\end{figure}

Preamplifier signals from the \Het proportional counters are analyzed by extracting the 10\%-50\% risetime and the maximum amplitude, which can be converted to energy deposited. 
The time location of the \Het signal is taken to be the half-height of the leading edge of the signal.
The risetime of \Het preamplifier signals has been previously shown to allow for particle identification to reject non-neutron capture events~\cite{2013LangfordHe3}. 
Figure~\ref{fig:HePlots} shows a scatter plot of the \Het risetime versus energy where different particle interactions have been identified. 
Also shown are microdischarge events, which are small sparks that produce very fast risetimes. 
 
The \Het detectors used in this work have an energy resolution of $\sim$2\% at the (n,$^3$He) capture peak, as is shown in the energy spectrum in Figure~\ref{fig:HePlots}. 
Approximately 20\% of neutron captures on \Het occur close enough to the detector walls such that either the proton or triton interact in the wall. 
This ``wall-effect'' is well-known and generates a low-energy tail in the detected energy spectrum.
The tail contains two discrete edges at 0.2~MeV and 0.6~MeV corresponding to either the proton or triton depositing its full energy in the wall~\cite{shalev1969,dietz1993,Amsbaugh2007}. 
For the work shown here, energy cuts of (0.2~MeV, 0.8~MeV) and risetime cuts of (0.1~$\mu$s, 0.7~$\mu$s) were made to maximize neutron efficiency while eliminating microdischarges and beta events.
As mentioned in Section~\ref{sec:detectorDesign}, alpha events were minimized by screening detectors and selecting those with minimal internal alpha activity.

\subsection{Coincidence analysis}

Fast neutrons events have a definite time-ordering (interaction in plastic scintillator followed by capture in \Het detector) and a characteristic time separation, dominated by neutron diffusion in the scintillator, that is well described by an exponential decay.  
Uncorrelated coincidences from gamma-ray interactions and thermal neutron captures will have no preferential time ordering and will have a uniform distribution of time separations.
In addition to being a neutron tag, the neutron capture signal indicates that the incident neutron fully thermalized in the detector. 
Due to their decreasing capture cross-section, neutrons that deposit only part of their energy in the detector have a small probability of capture on \Het. 
Thus, the neutron spectrum detected with FaNS-2 only contains full-energy depositions, which provide better spectral information than partial depositions.  

\begin{figure}[h]
\begin{center}
\includegraphics[width=0.6\textwidth]{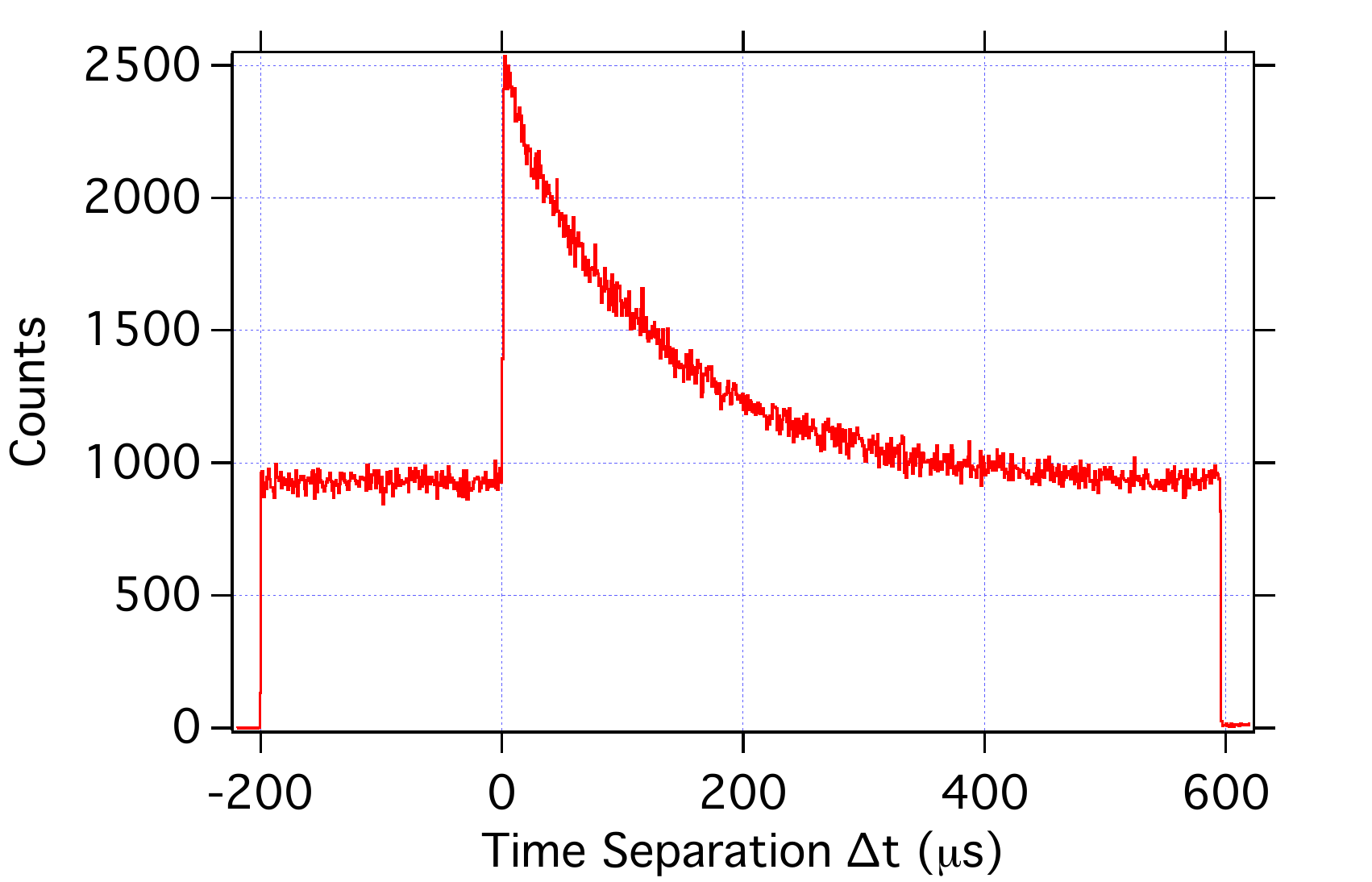}~
\caption{The time separation ($\Delta$t) between scintillator and \Het signals for FaNS-2. Negative time separations are non-physical events where the \Het signal occurs before the scintillator. 
These uncorrelated coincidences are subtracted from positive time separation events in analysis.}
\label{fig:timeSep}
\end{center}
\end{figure}

An example distribution of \Het - PMT time separations is shown in Figure~\ref{fig:timeSep}.
Here we define the time separation, $\Delta$T, as the difference between the 10\% height of the \Het signal and the 50\% height of the PMT signal cluster: $\Delta$T = T$_{\textrm{He}}$ -  T$_{\textrm{PMT}}$.
The positive time separation event distribution is fitted with a single exponential with 125~$\mu$s lifetime.
Therefore, a coincidence window of 600~$\mu$s is sufficient to detect greater than 99\% of all correlated captures.
Events where the \Het signal occurs before the scintillator, i.e. those with time separations between -200~$\mu$s and 0~$\mu$s, are used to directly monitor uncorrelated coincidences and measure their properties. 
The uncorrelated energy spectra are subtracted from those with positive time separation, producing background-subtracted neutron energy spectra. 
This process is demonstrated in Section~\ref{sec:monoEresponse}.

This coincidence window is sufficiently large to expect multiple clusters PMT signals in each acquisition window. 
Each PMT cluster is treated independently, and all clusters are considered as possible candidates for the neutron recoil signal. 
The example digitized acquisition window in Figure~\ref{fig:traces} shows two clusters, labeled C0 and C1, of multiple PMT signals. 
There is no way to discriminate on an event-by-event basis which of these is the true neutron recoil associated with the \Het signal.
However, it is possible to statistically subtract uncorrelated signals provided all clusters of PMT signals are included in the analysis.

\section{Characterization with calibrated sources}
\label{results}
\subsection{Response to gamma calibration sources}

\begin{figure}[h]
\begin{center}
\includegraphics[width=.48\textwidth, height=1.8in]{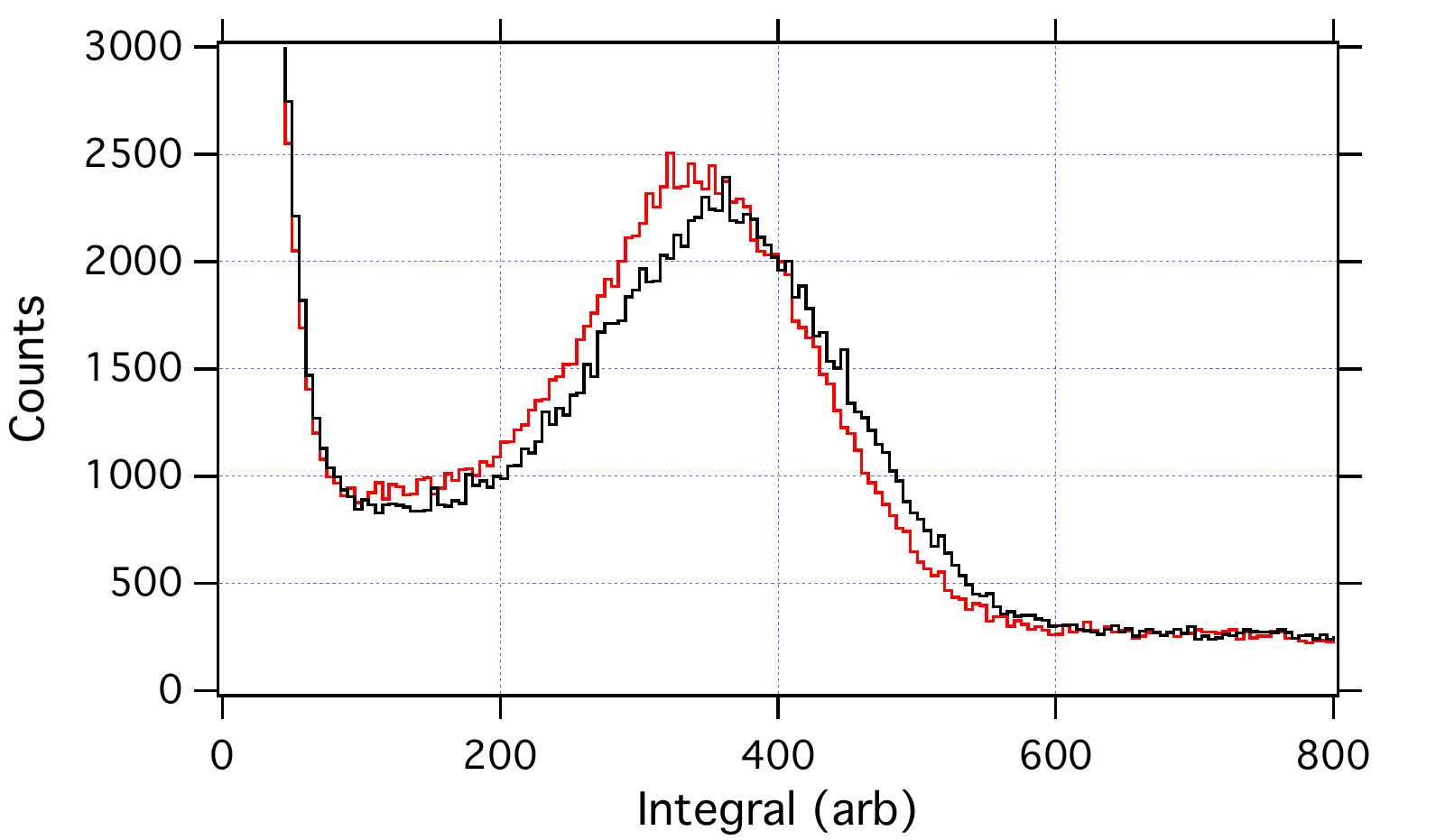}~
\includegraphics[width=.48\textwidth, height=1.8in]{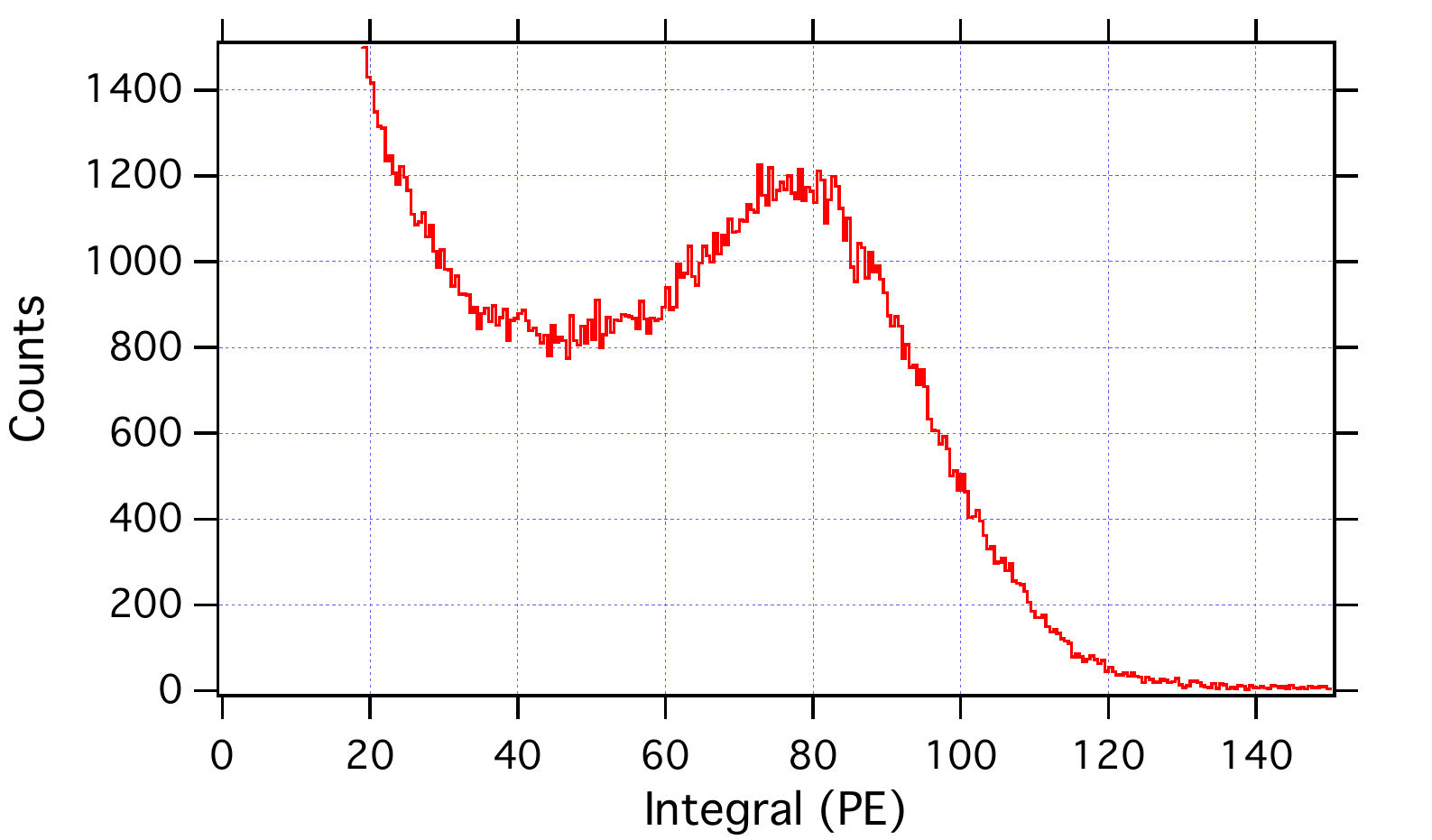}
\caption{Left: Single PE distributions for two PMTs on one of the FaNS-2 scintillator segments. Right: The combined PE spectrum from the PMTs when exposed to a \Cs source.}
\label{fig:CsPeMeV}
\end{center}
\end{figure}

To determine the light collection efficiency, the single photoelectron responses were measured for each PMT, examples of which are shown in Figure~\ref{fig:CsPeMeV}. 
Then a series of measurements were made with gamma calibration sources, specifically \Cs and \Co. 
The gamma interactions are dominated by single Compton scattering, which result in characteristic spectra with half-heights of approximately 0.48~MeV and 1.0~MeV respectively. 
The detected spectra were compared to simulation and the PE/MeV conversion extracted.
An example of the \Cs spectrum is shown in Figure~\ref{fig:CsPeMeV}.
The combined light collection efficiency of a single FaNS-2 scintillator segment is determined to be $\sim$200~PE/MeV.

During operation, periodic calibrations are performed using the ambient gamma ray and muon backgrounds. 
Most prominent gamma lines are from $^{40}$K and $^{208}$Tl at 1.46~MeV and 2.6~MeV respectively. 
The individual segments are 9~cm thick, which yield a muon minimum-ionizing peak of 18~MeV.
These provide a range of calibration points that allow for tracking any changes of PMT gain. 

\subsection{Absolute neutron efficiency} 
FaNS-2  operated in the Low Scatter Facility at the National Institute for Standards and Technology in Gaithersburg, MD. 
The Low Scatter Facility is a large-volume room specially constructed for neutron calibrations using low mass walls to minimize the back scattering  of neutrons that may return to the detector, often referred to as ``room-return''.
To determine the absolute neutron detection efficiency, a calibrated $^{252}$Cf spontaneous fission source was deployed at multiple distances from the detector.
The source, DHS-9667, is enclosed in a $\sim3$~cm diameter steel housing and had a total neutron emission rate of (5690$\pm$150)~/s when these data were collected, of which (2137$\pm$85)~/s are above an analysis threshold of 2~MeV$_n$~\cite{2009rdsa.conf..361G}.
The source was deployed at seven distances from the top of the detector, ranging from 85~cm to 238~cm.

Because FaNS-2 is not small compared to the distances at which the source was positioned, a $1/r^2$ estimate of the solid angle cannot be used. 
The fractional solid angle, $\Omega$, has been calculated for each source position using a two-dimensional integral

\begin{equation}
\Omega = \int_{SA} \frac{h+z_0}{(x^2 + y^2 + (h+z_0)^2)^{3/2}}\, dx\, dy,
\label{FaNS2solidAngleEq}
\end{equation}

\noindent where $x$ and $y$ are taken to be in the plane of the surface of the detector, $z_0$ is the average interaction depth of neutrons in the detector, and $h$ is the distance above the top of the detector where the source was positioned, with integration limits covering the top plane of the detector. 
The distances are measured from the bottom of the source enclosure to the center of the first layer of scintillator, which is the approximate depth of interaction for \Cf energy neutrons~\cite{Pozzi2004}. 
An MCNP simulation was performed for each source position, with detected neutron count rates determined by applying analysis cuts to simulated data. 
As discussed in Section~\ref{sec:detectorDesign}, a \Het detector efficiency of (84$\pm$10)\% was applied to the MCNP data to account for the overproduction of neutron captures in the simulations. 
During the \Cf data collection, two long data sets were run without the source. 
These rates have been subtracted from each source position to isolate the rate due to the \Cf source.

\begin{figure}[h]
\begin{center}
\includegraphics[width=0.6\textwidth]{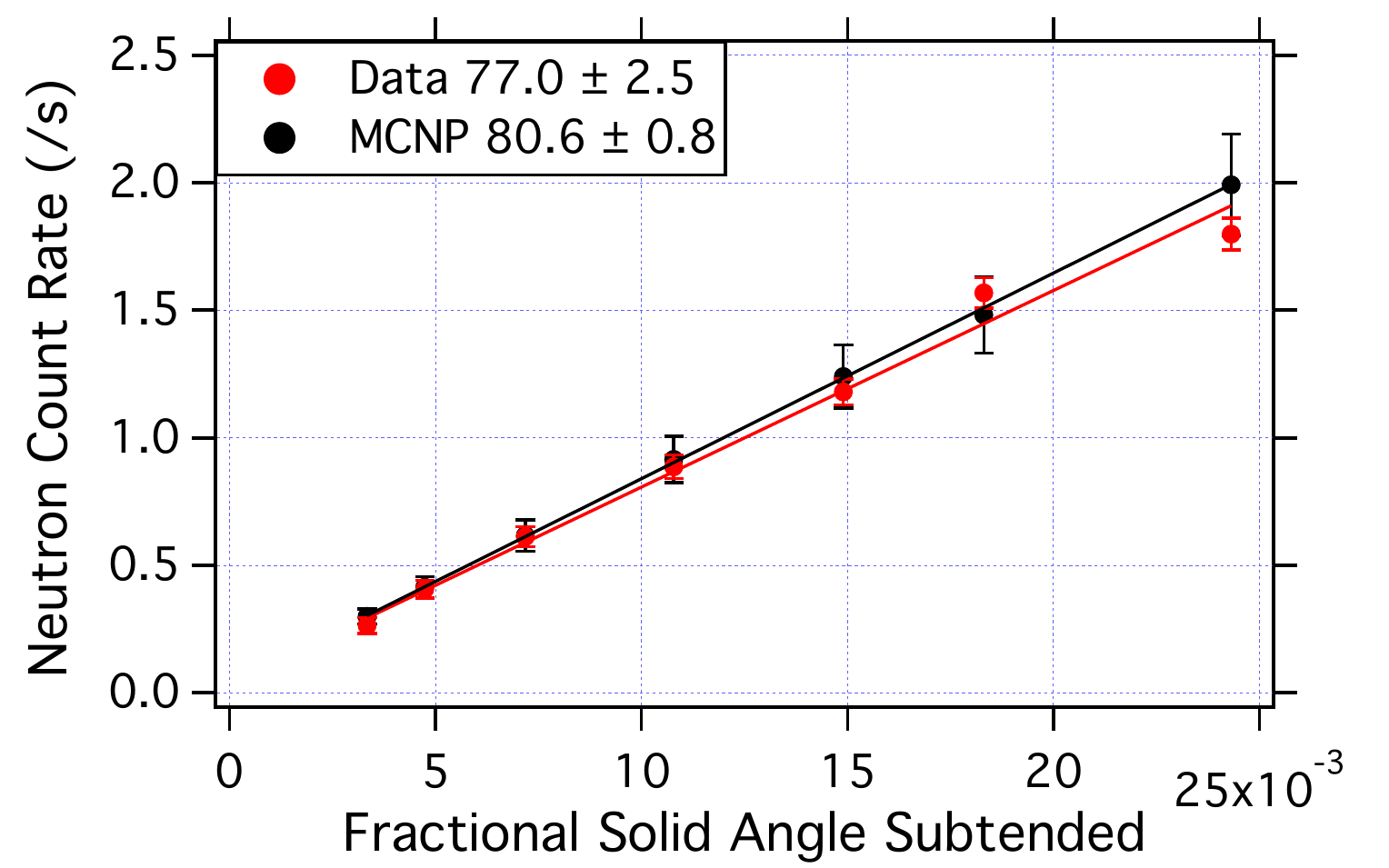}
\caption{The resulting detected neutron rate (after subtracting the ambient neutron rate) versus subtended solid angle for a \Cf source at multiple distances (red) with statistical error bars. The MCNP predictions for each distance (black) are shown after the \Het detection efficiency of 84\% has been applied. The slope of each linear fit is listed in the legend.}
\label{fig:FaNS2Eff20MeV}
\end{center}
\end{figure}

As demonstrated in Figure~\ref{fig:FaNS2Eff20MeV}, there is very good agreement between the data and MCNP observed in these measurements. 
The fitted slope of these data is proportional to the efficiency divided by the source activity
\begin{equation}
\varepsilon_{E_n > 2~MeV} = \frac{slope}{\Gamma_s} = \frac{(77\pm2.5)~\textrm{/s}}{(2137 \pm85)~\textrm{/s}} = (3.6\pm 0.15)\%.
\end{equation}

\noindent The performance of the simulation in reproducing the detected efficiency demonstrates that FaNS-2 is well-modeled by MCNP.
Therefore, one would reasonably expect that simulations of the detector response to other sources with similar energy ranges should also match experimental results.
This would allow FaNS-2 to be used to study and characterize the flux and spectra from a variety of neutron sources such as Am-Be or Pu-Be.

\subsection{Response to monoenergetic neutron generators}
\label{sec:monoEresponse}
\begin{figure}[h]
\begin{center}
\includegraphics[width=0.48\textwidth, height=1.9in]{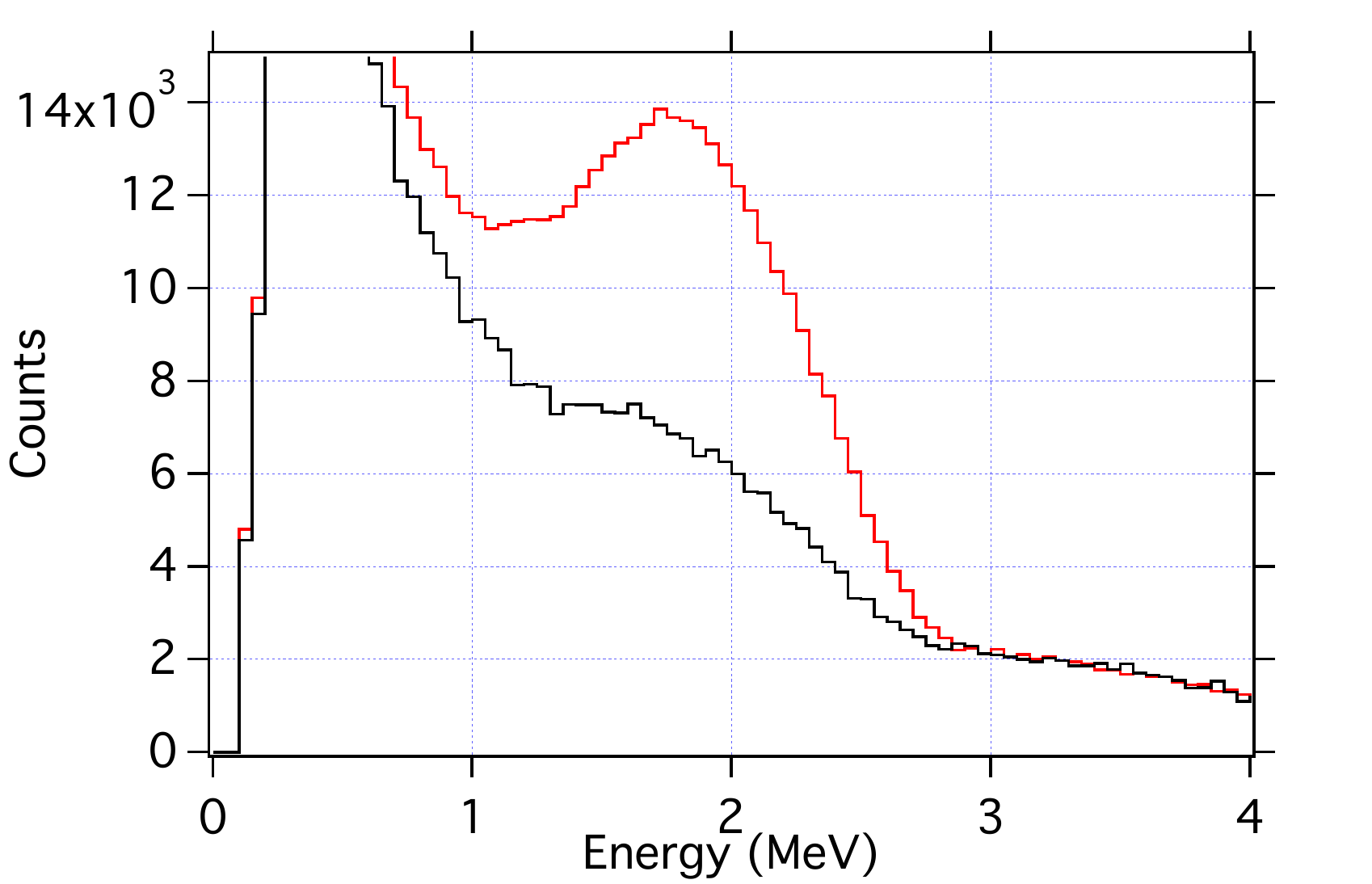}~
\includegraphics[width=0.48\textwidth, height=1.9in]{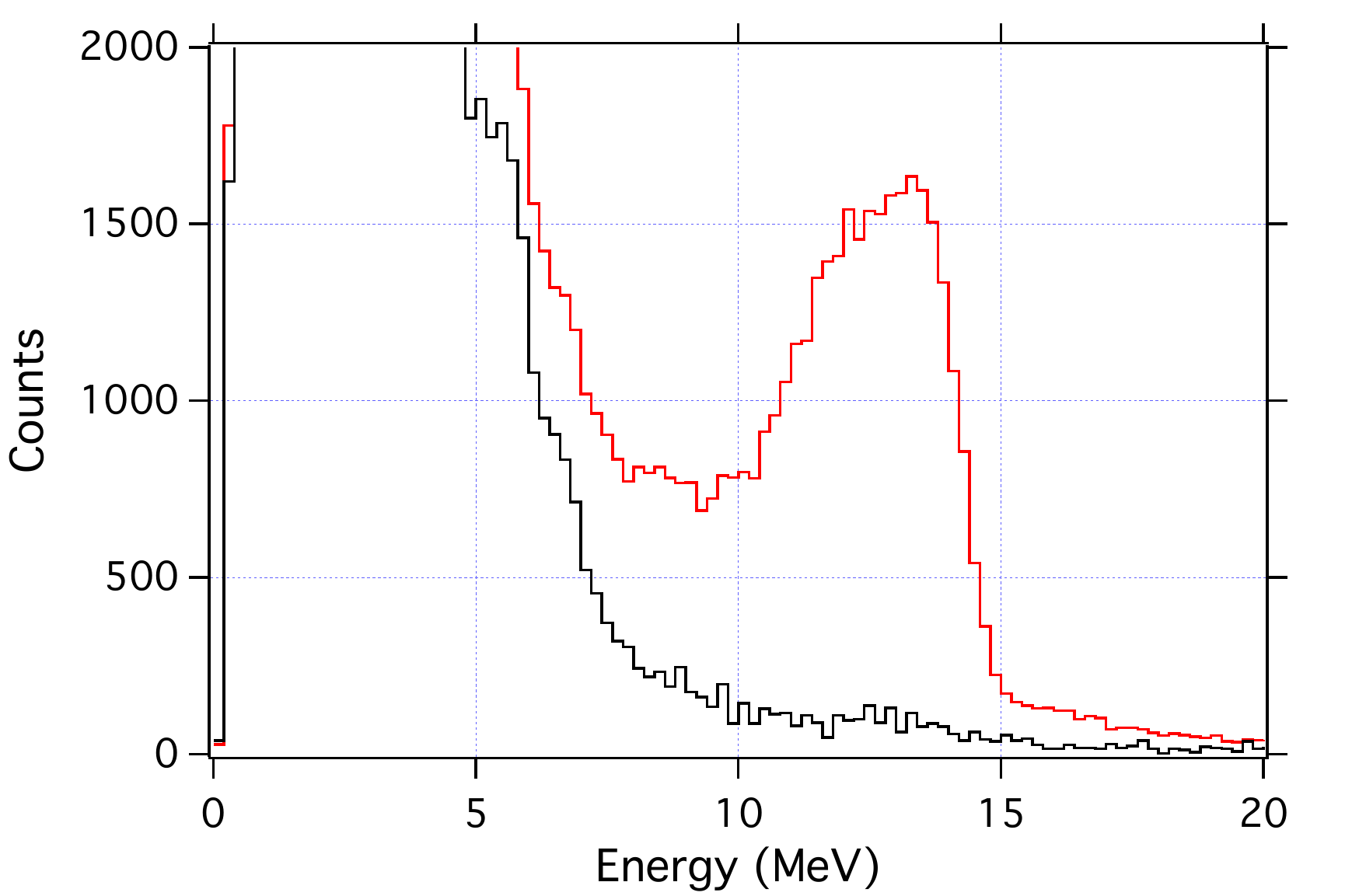}\\
\includegraphics[width=0.48\textwidth, height=1.9in]{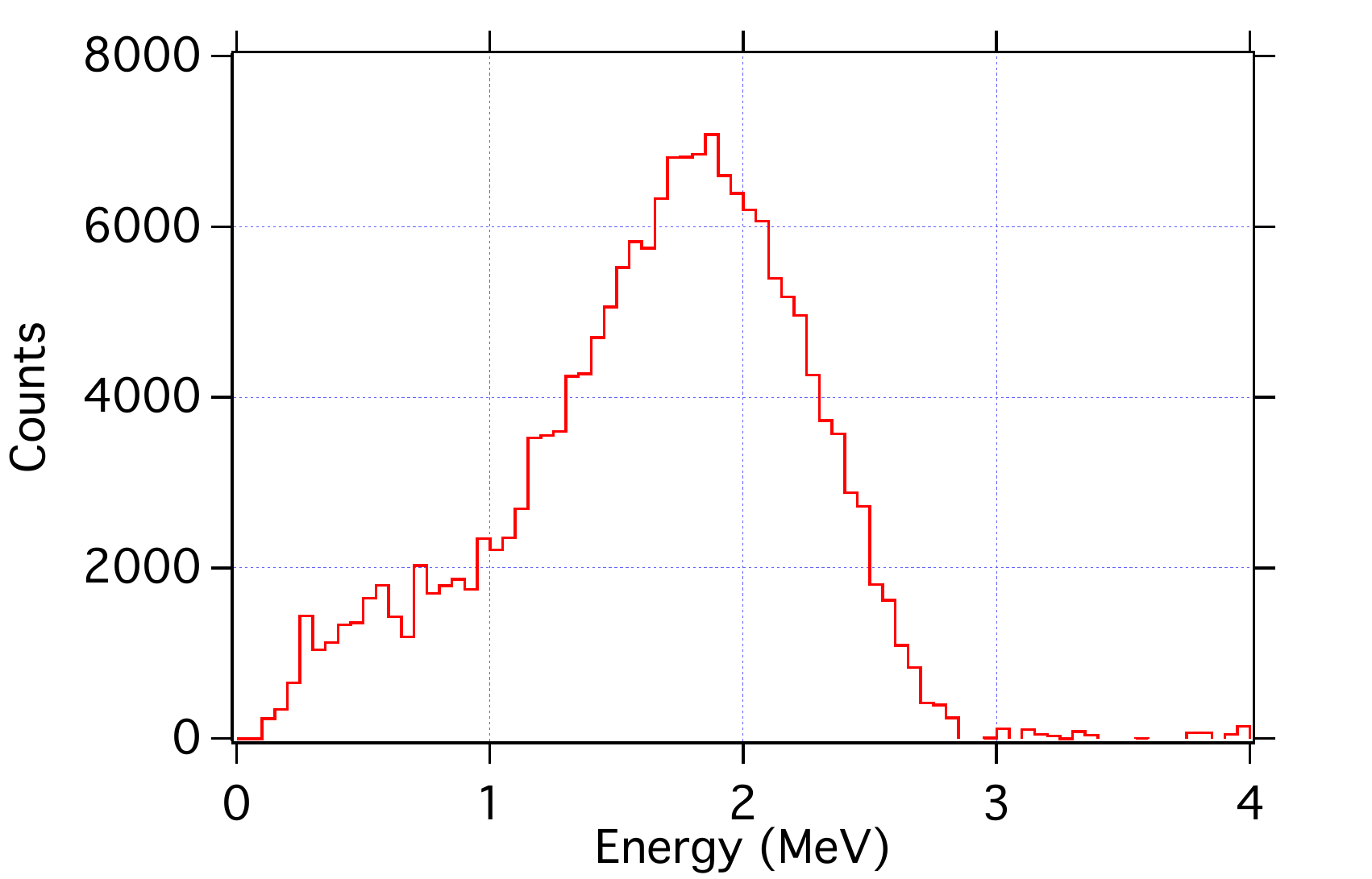}~
\includegraphics[width=0.48\textwidth, height=1.9in]{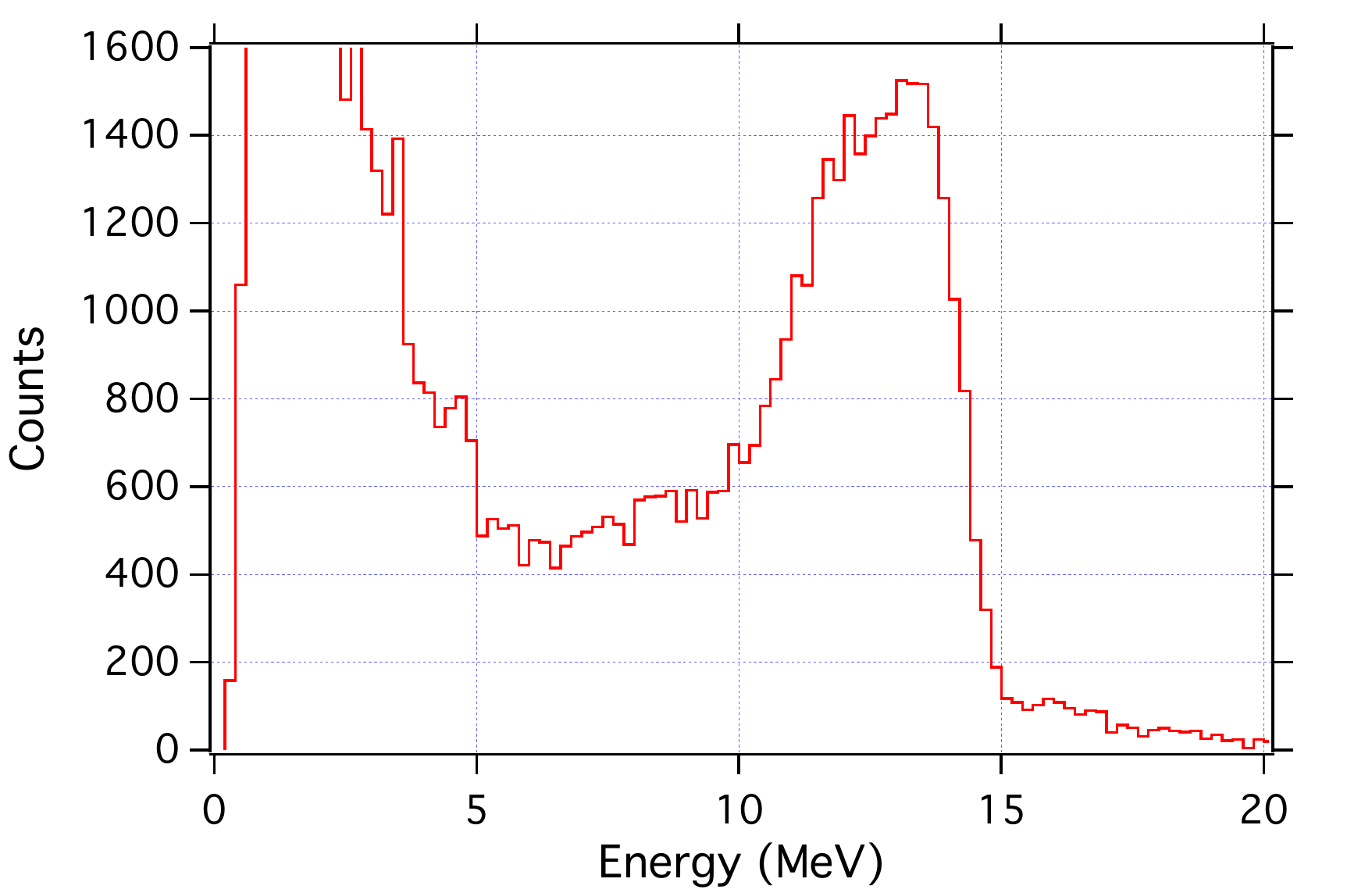}
\caption{Top: Reconstructed energy spectra for the DD (left) and DT (right) monoenergetic neutron generators. The red spectra are correlated coincidences, while the black spectra are uncorrelated coincidences after being scaled to account for the asymmetric coincidence window. The uncorrelated events are dominated by generator neutrons that do not fully thermalize in the detector. Bottom: Energy spectra for the DD (left) and DT (right) monoenergetic neutron generators after subtraction of uncorrelated coincidences. The widths of the peaks are enhanced by neutron multiple scattering in a single segment that degrades the reconstruction.}
\label{fig:monoEresults}
\end{center}
\end{figure}

Two monoenergetic neutron generators based on deuterium-deuterium (DD) and deuterium-tritium (DT) fusion were utilized for characterization of FaNS-2.
The generators produce neutrons at 2.5~MeV and 14~MeV that are excellent for demonstrating the energy response of FaNS-2.
The generators were positioned directly above the center of the detector at approximately 20~cm from the first layer of scintillator.
The close source proximity and operation in the Low Scatter room minimize room-return neutrons that would enter the detector with lower energy than the primary neutron flux.

Figure~\ref{fig:monoEresults} show the correlated and uncorrelated energy spectra for 2.5 and 14~MeV neutrons after scaling the negative time separations to account for different acceptance window. 
The uncorrelated signals are dominated by generator neutrons that do not fully thermalize in the detector. 
After subtracting the uncorrelated events from the signal coincidences, the resulting reconstructed energy spectra are shown in the bottom panels of Figure~\ref{fig:monoEresults}. 
Well-defined peaks are visible for both monoenergetic neutron sources, and the upper edges of the reconstructed spectra correspond to the maximum incident neutron energy, 2.5~MeV or 14~MeV.

\subsection{Energy resolution through segmentation}
\label{sec:segmentation}
\begin{figure}[h]
\begin{center}
\includegraphics[width=.55\textwidth]{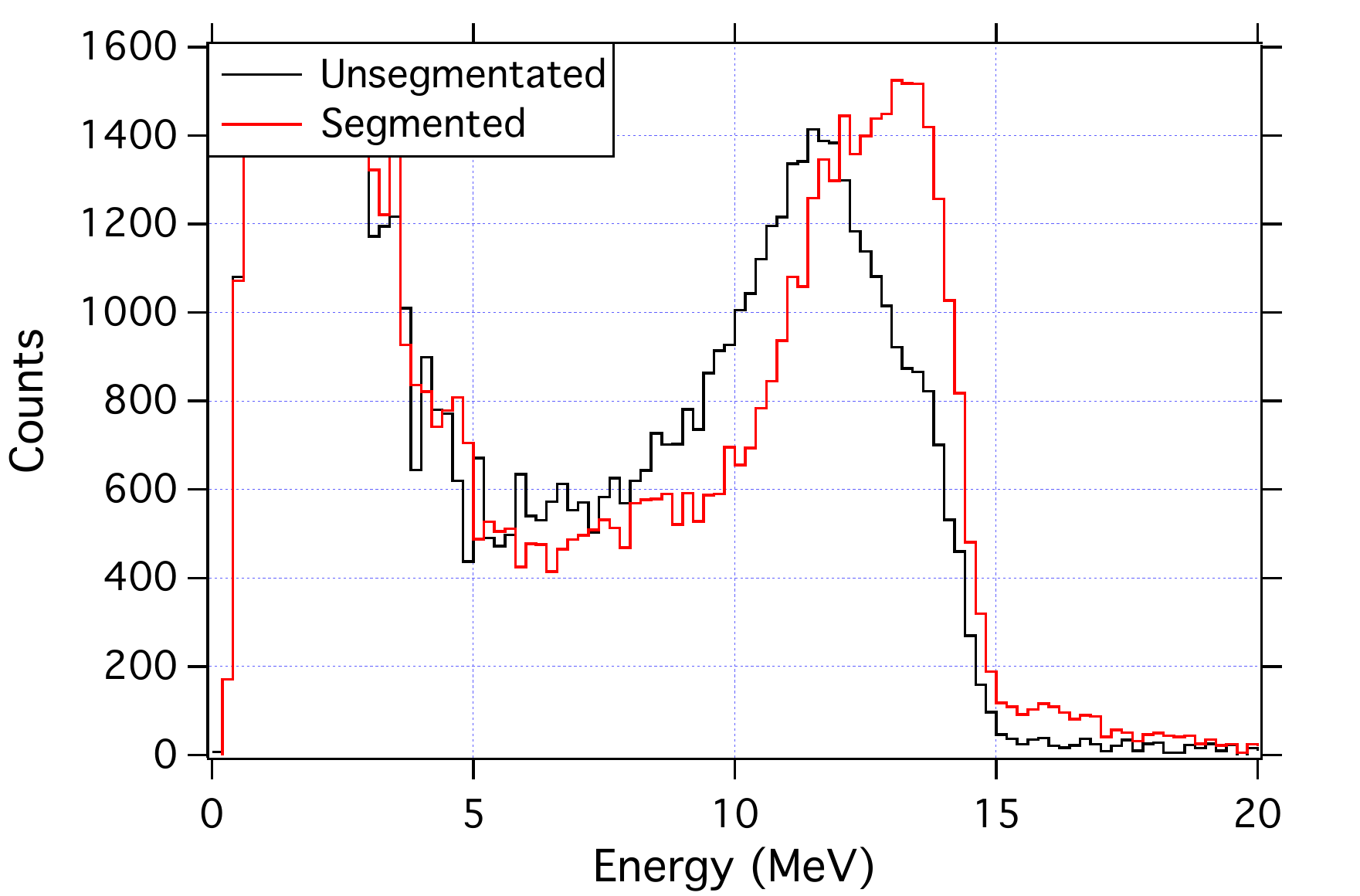}
\caption{Comparison between DT energy spectra determined by treating all of FaNS-2 as a single scintillation volume or using the segmentation to remove nonlinear effects.}
\label{fig:segmentation}
\end{center}
\end{figure}

Because energy depositions in multiple scintillator segments can be reconstructed separately, the segmented nature of FaNS-2 allows for improved energy reconstruction over monolithic volume detectors. 
To directly demonstrate this improvement the 14~MeV energy spectrum can also be reconstructed as if FaNS-2 were one volume. 
Figure~\ref{fig:segmentation} shows the comparison of the energy spectra from the monolithic analysis and from utilizing the segmentation.  

The unsegmented energy spectrum does not produce the expected full-energy of 14~MeV for most events.
The main peak is at approximately 12~MeV with a slight shoulder at $\sim$14~MeV from events that deposit all their energy in a single scatter.
The segmented reconstruction produces a significantly improved spectrum with a narrower peak and a sharp cutoff at 14~MeV.
Even though the segments in FaNS-2 are relatively large, the observed improvement highlights the benefits of a segmented detector.  
Finer segmentation would reduce the inflection in the reconstructed spectra at $\sim$12~MeV that arises from neutrons that scatter twice in one segment.

\subsection{Singular Value Decomposition unfolding}
\label{sec:unfolding}
\begin{figure}[h]
\begin{center}
\includegraphics[width=0.48\textwidth, height=2.in]{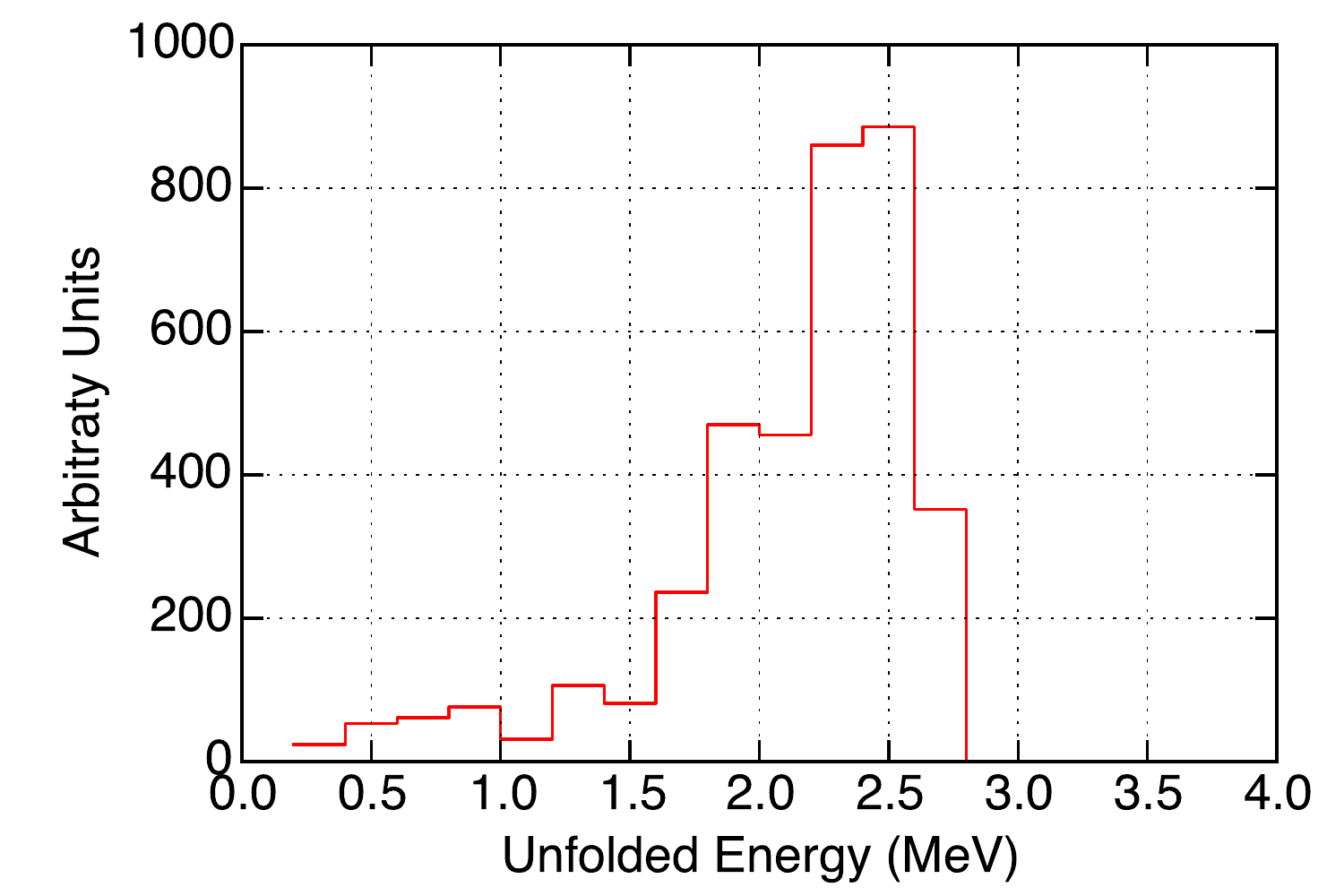}
\includegraphics[width=0.48\textwidth, height=2.in]{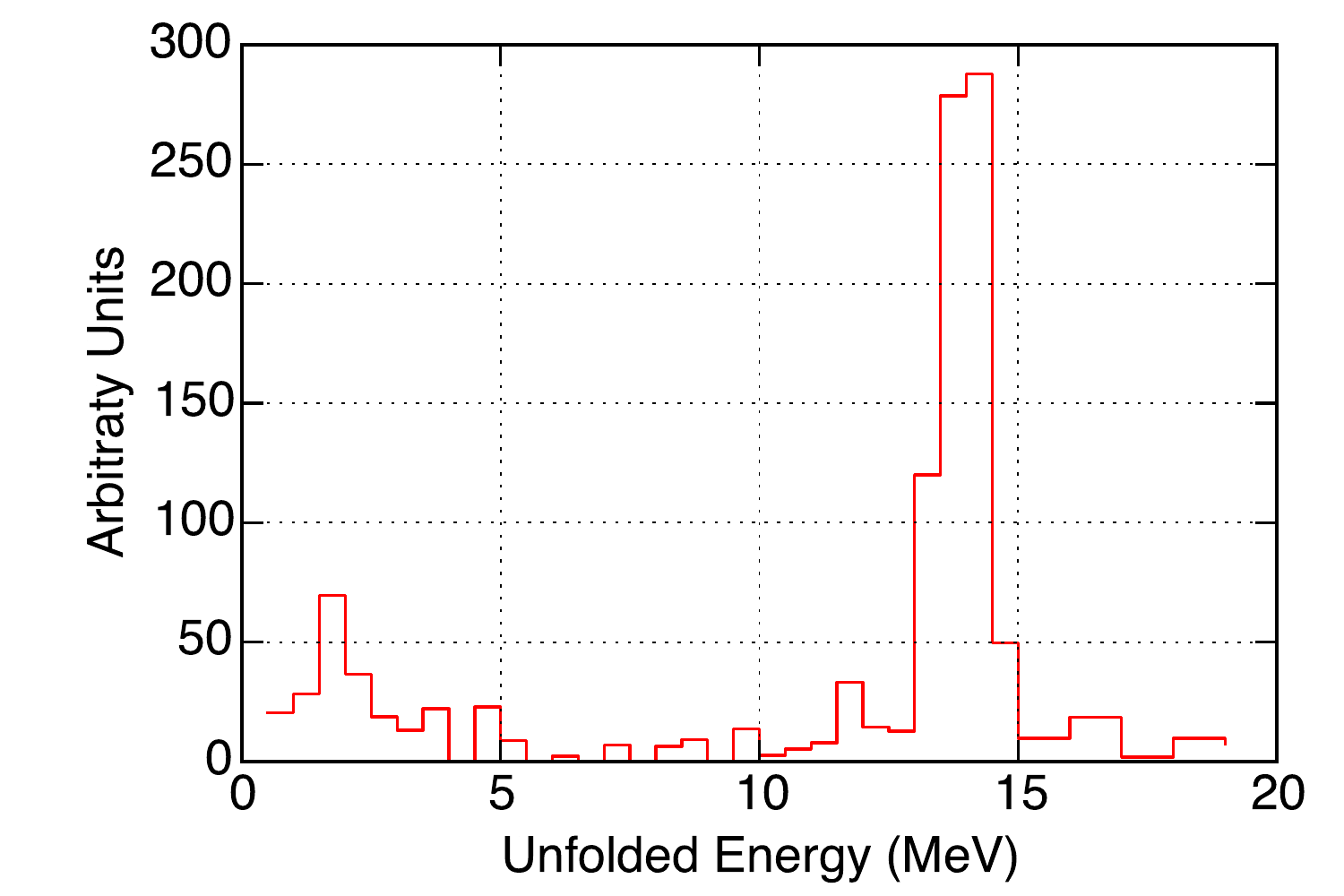}
\caption[]{Unfolded energy spectra for the DD (left) and DT (right) monoenergetic neutron sources. Improved energy resolution compared to Figure~\ref{fig:monoEresults} is observed in both spectra.}
\label{fig:unfolded}
\end{center}
\end{figure}

In addition to the improvements made via capture-gated spectroscopy, further gains in energy response are still available through unfolding.
Detector response functions based on MCNP simulations discussed earlier can be used to form a response matrix for Singular Value Decomposition (SVD).
This technique has been described at length elsewhere; detailed discussions may be found in Refs.~\cite{hoecker1996svd,2011Adye, 2011Tackmann}. 
Coarsely binned 2.5 and 14~MeV neutron energy spectra are passed through an SVD routine, using a cutoff of 0.001 times the highest singular value. 
The resulting output spectra are shown in Figure~\ref{fig:unfolded}.
Both spectra show significant improvements in peak-widths and the peaks are now centered at the expected energy of the respective monoenergetic neutron sources.
The 14~MeV neutron spectrum, Figure~\ref{fig:monoEresults}, before unfolding has a significant fraction of events at lower energy, which Monte Carlo studies indicate are a result of neutron interactions on carbon. 
By fitting a Gaussian to the peak region of the 14~MeV spectrum, we observe an energy resolution of $\sim$5\%. 
Further refinements of the SVD unfolding procedure can be undertaken to enhance these results, but this initial unfolding demonstrates the possible gains from unfolding. 

\subsection{Directionality of detector response and depth of interaction}
\label{sec:directionality}

\begin{figure}[h]
\begin{center}
\includegraphics[width=0.32\textwidth, height=1.56in, trim={.2in .2in .2in .2in},clip]{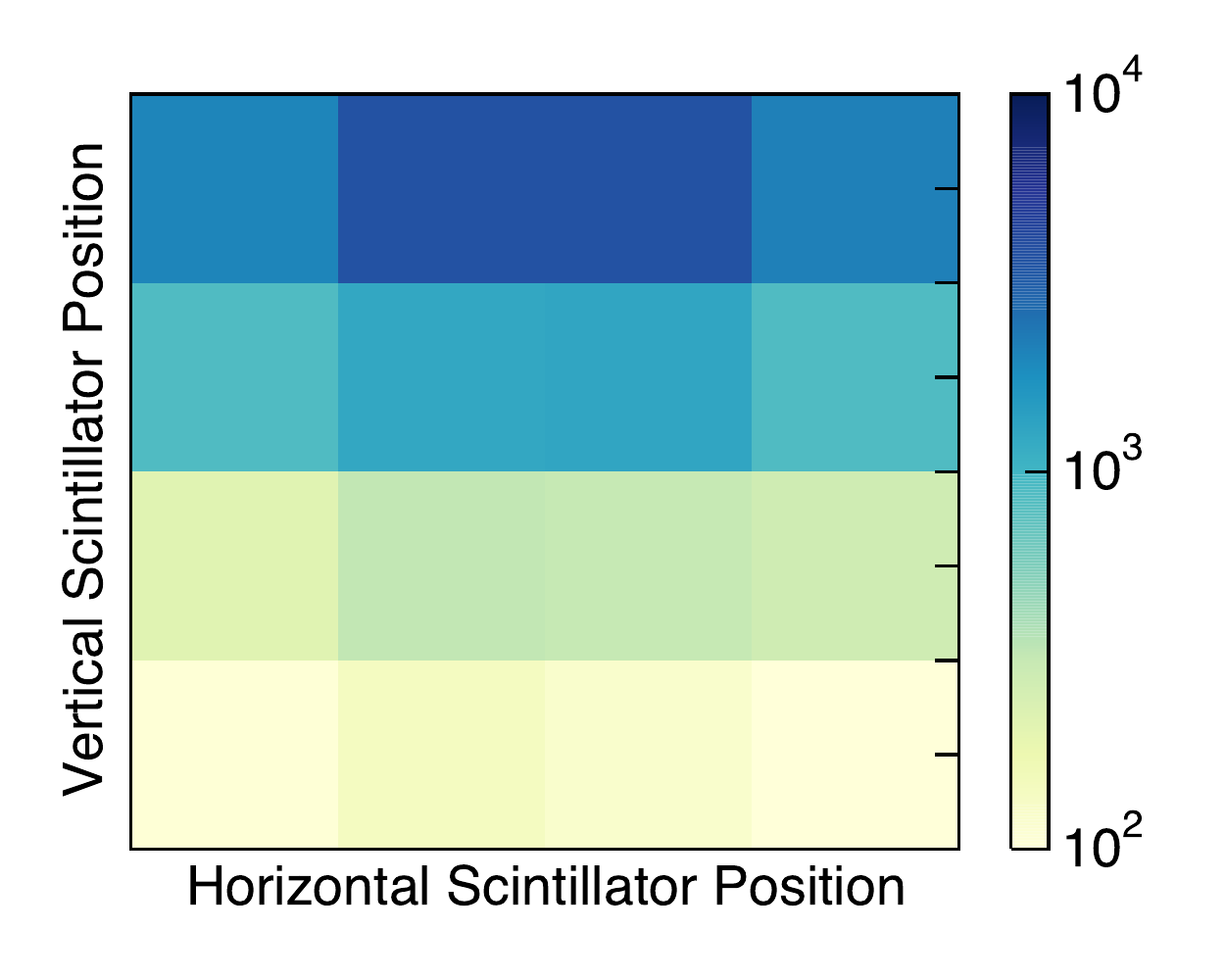}~
\includegraphics[width=0.32\textwidth, height=1.56in, trim={.2in .2in .2in .2in},clip]{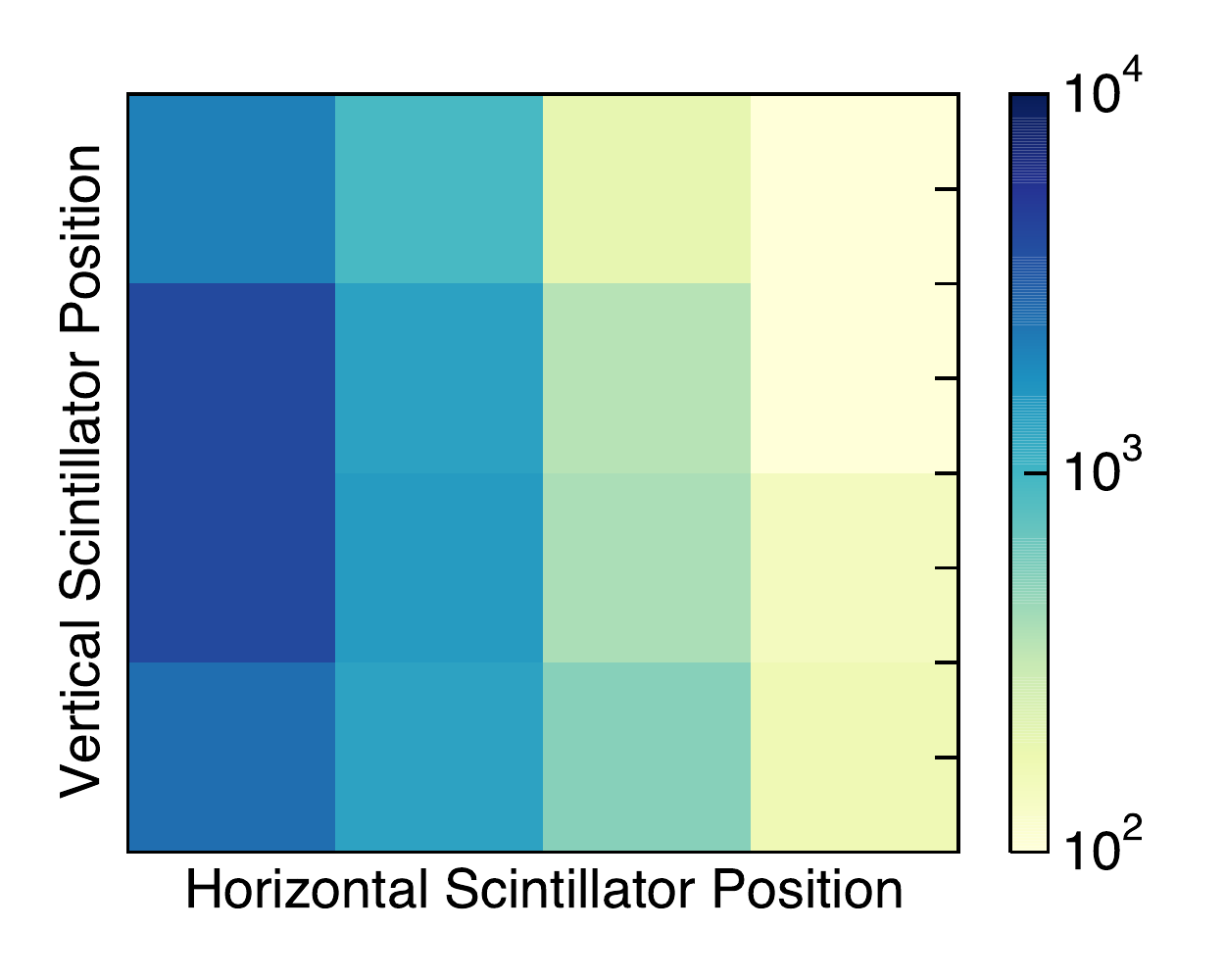}~
\includegraphics[width=0.32\textwidth, height=1.56in, trim={.2in .2in .2in .2in},clip]{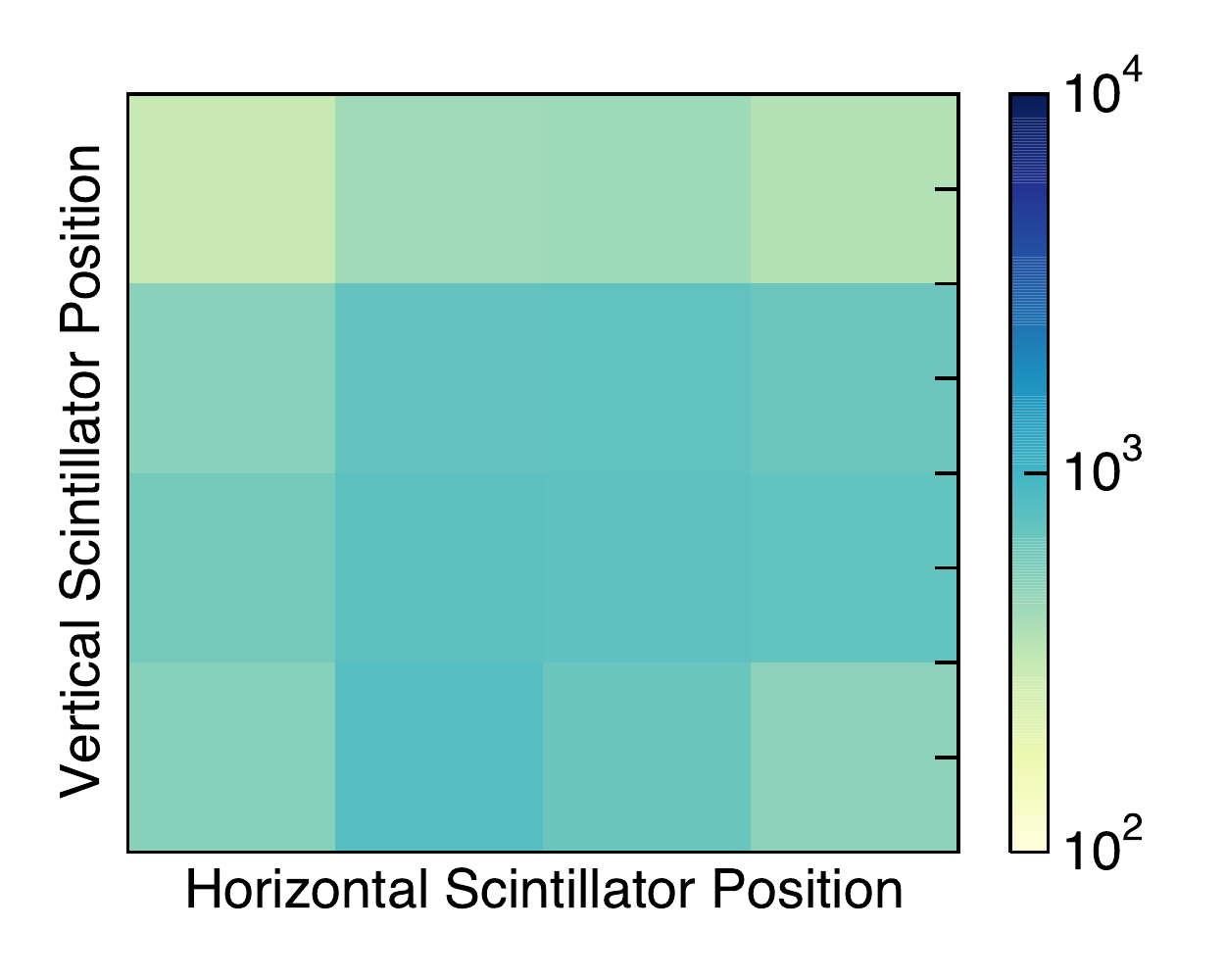}
\caption{Histograms of neutron scatter locations in the scintillator segments (displayed in a grid similar to the detector layout) from data collected with the \Cf source 50~cm above (left), to the side (center), and end-on (right) after subtracting ambient backgrounds. Enhanced interaction rates can been seen in the top and side layers for the left and center figures. The right figure shows how an end-on source illuminates the detector symmetrically.}
\label{fig:heatplots}
\end{center}
\end{figure}

As a segmented detector, the direction of a neutron source can be identified by comparing the rate of interactions in each detector element.
Figure~\ref{fig:heatplots} shows the distribution of neutron scatter locations in the scintillator segments from data collected with the \Cf source positioned 50~cm above, to the side, and along the detector's axis of symmetry, see Figure~\ref{fig:FaNS2diagram}.
For sources placed above and to the side of the detector, we observe an asymmetric enhancement of counts in detectors close to the source position. 
For a source deployed end-on, i.e. along the detector's axis of symmetry, all detectors are illuminated equally and we observe a symmetric distribution of neutron interactions.
Refinement of the position determination can be made through combining scintillator and \Het detector information. 
Directionality may be particularly useful in applications such as the detection of illicit neutron sources or special nuclear material~\cite{5076056}.

\begin{figure}[h]
\begin{center}
\includegraphics[width=0.6\textwidth]{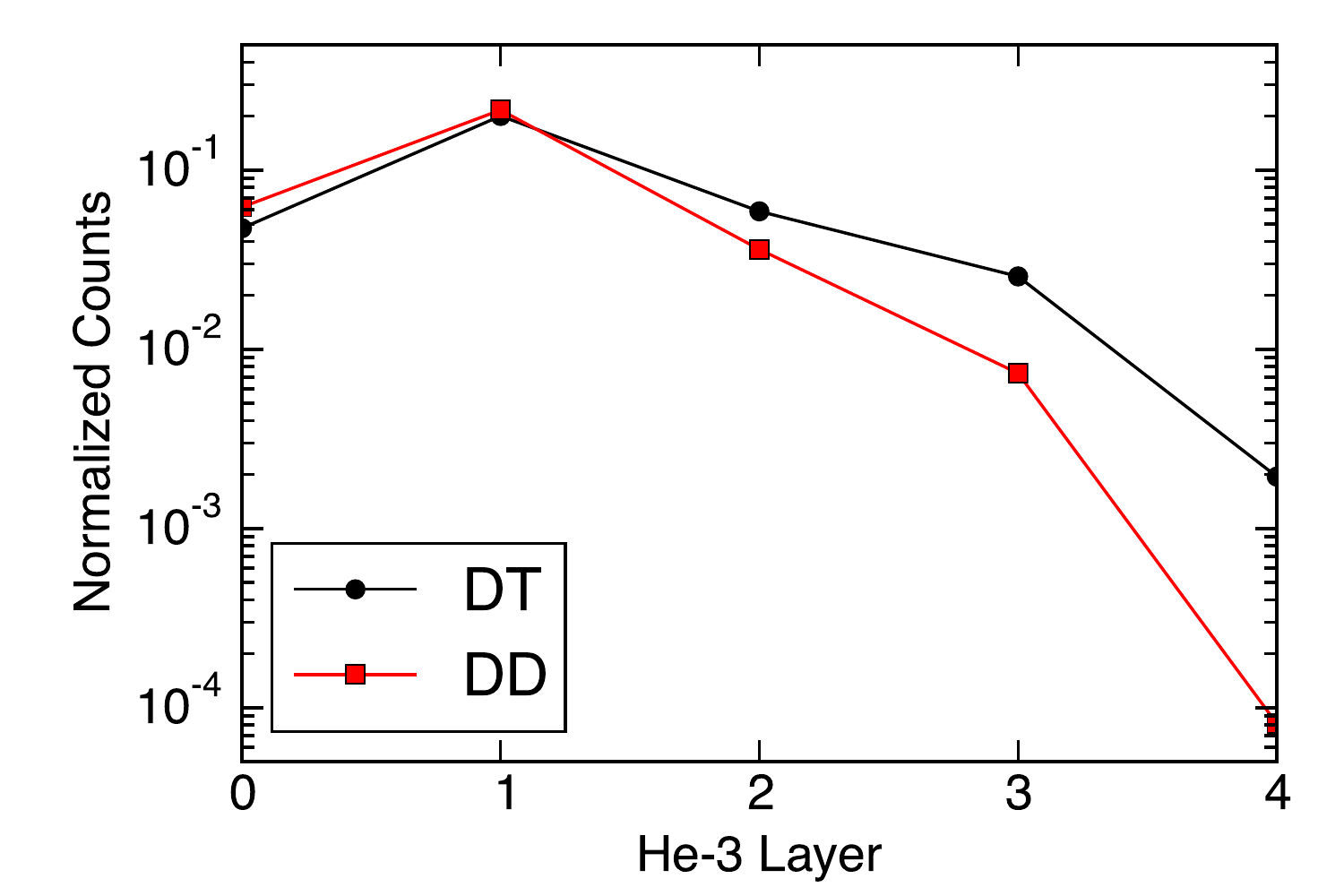}~
\caption{Normalized distribution of \Het neutron capture location within the detector from data collected with the 2.5~MeV (red) and 14~MeV (black) neutron generators after subtracting ambient backgrounds. Layer-0 is the top of the detector, while Layer-4 is the bottom. The higher energy neutrons have more deeply penetrating neutron captures. }
\label{fig:depthDist}
\end{center}
\end{figure}

Higher energy neutrons are more deeply penetrating in FaNS-2, which can be used to further identify source location and type. 
To demonstrate this, we have compared the depth of neutron capture for the DD and DT monoenergetic neutron generators. 
Both generators were placed at 20~cm above FaNS-2 and operated at a similar intensity.
Figure~\ref{fig:depthDist} shows the distribution of neutron captures as a function of \Het layer.   
We observe that the 14~MeV neutrons capture on \Het detectors deeper in the detector, as expected.
The average capture depth of 2.5~MeV (14~MeV) neutrons is 0.97 (1.2) layers. 
Identification of neutron source type can be enhanced by combining energy spectrum information with the average depth of interaction.  
Furthermore, it may be possible to distinguish multiple neutron sources with different energy spectra using position and energy information. 

\section{Summary and outlook}

An improved high-sensitivity fast neutron detector, FaNS-2, has been developed and characterized with multiple neutron sources.
A neutron time-of-flight measurement has characterized the nonlinear light response of the FaNS-2 scintillator, allowing for precise reconstruction of neutron events with multiple scatters. 
Using a calibrated \Cf neutron source, the absolute neutron detection efficiency of FaNS-2 is determined to be (3.6$\pm$0.15)\% for neutrons with energies above 2~MeV. 
Through segmentation and capture-gating, FaNS-2 displays an excellent response to both 2.5 and 14~MeV monoenergetic neutrons without unfolding. 
After implementing an initial unfolding technique using Single Value Decomposition, FaNS-2 has a demonstrated $\sim$5\% energy resolution at 14~MeV.
Further refinement of the unfolding could yield even improved energy resolution. 

FaNS-2 has now been deployed at the earth's surface in a low-shielding environment and in a shallow underground lab ($\sim$20~meter water equivalent) at NIST to study the primary and secondary cosmogenic neutron spectra and fluxes.
At the surface, high energy neutrons are produced in the upper atmosphere in extended air-showers. 
These neutrons range in energy from below 1~MeV to many TeV, with fluxes that vary by orders of magnitude~\cite{Gordon2004}. 
At $\sim$20~meter water equivalent depth, the majority of air-shower neutrons have been attenuated leaving only those produced by cosmic-ray interactions in the material overhead~\cite{aguayo2011cosmic}. 
By measuring the spectra in both environments with the same detector, a relative comparison can be made and compared to simulations with reduced systematic uncertainties.
Simulations indicate that FaNS-2 has approximately an order of magnitude higher sensitivity to cosmogenic neutrons than FaNS-1.  
The enhanced sensitivity extends the detectors capability to characterize low fluxes of neutrons over a wide range of energies and in the presence gamma backgrounds. 

\acknowledgments

We acknowledge the NIST Center for Neutron Research for the loan of the \Het proportional counters used in this work. 
The research has been partially supported by NSF Grant 0809696. 
T.J.~Langford acknowledges support under the National Institute for Standards and Technology American Recovery and Reinvestment Act Measurement Science and Engineering Fellowship Program Award 70NANB10H026 through the University of Maryland

\bibliographystyle{h-physrev}
\bibliography{main} 

\end{document}